\documentclass[prd,tightenlines,floatfix,showpacs,preprintnumbers,nofootinbib,eqsecnum]{revtex4}
 \usepackage[dvips,final]{graphicx}
  \usepackage{amssymb}
   \usepackage{amsmath} % Advanced math typesetting
    \usepackage{amsfonts}
     \usepackage{epsfig}
     \usepackage{color}
      \usepackage{bm} % bold math

\usepackage{booktabs}
\usepackage{multirow}
\usepackage{slashed}
\usepackage{comment}

\usepackage{cancel}
\usepackage[normalem]{ulem}

\usepackage{amscd}
\usepackage{subfigure}
\def\doublespace{\def\baselinestretch{1.6}\large\normalsize}
\def\normalspace{\def\baselinestretch{1.0}\normalsize}
\def\PSfig#1#2{\scalebox{#1}{\includegraphics{#2}}}

\def\Caption#1{
  \normalspace
  \vskip-1mm\caption{\sl#1}\vskip-1mm
  \doublespace
}

\def\BA{\begin{eqnarray}}
\def\BE{\begin{equation}}
\def\BF{\begin{figure}[htb]}
\def\BT{\begin{table}[htb]}
\def\EA{\end{eqnarray}}
\def\EE{\end{equation}}
\def\EF{\end{figure}}
\def\ET{\end{table}}
\def\la{\langle}
\def\ra{\rangle}

\def\TeV{\,\mbox{TeV}}
\def\GeV{\,\mbox{GeV}}

\def\Jpsi{J\!/\!\psi}
\def\psip{\psi^{\,\prime}}

\def\sqq{\sigma_{Q\bar Q}}

\def\aem{\alpha_{em}}

\def\lsim{\mathrel{\rlap{\lower4pt\hbox{\hskip1pt$\sim$}}
     \raise1pt\hbox{$<$}}}         %less than or approx. symbol
 \def\gsim{\mathrel{\rlap{\lower4pt\hbox{\hskip1pt$\sim$}}
     \raise1pt\hbox{$>$}}}         %greater than or approx. symbol
\begin{document}
%%%%%%%%%%%%%%%%%

%================================================
\title{
Coherent {photo- and}
electroproduction of charmonia on nuclear targets revisited:
\\
Green function formalism
\vspace*{0.20cm}
}
%================================================

\author{J. Nemchik$^{1,2}$}
\email{jan.nemcik@fjfi.cvut.cz}

\author{J. \'Obertov\'a$^{1}$}
\email{jaroslava.obertova@fjfi.cvut.cz}

\affiliation{
\vspace*{0.20cm}
$^1$
Czech Technical University in Prague, FNSPE, B\v rehov\'a 7, 11519
Prague, Czech Republic}
\affiliation{$^2$
Institute of Experimental Physics SAS, Watsonova 47, 04001 Ko\v sice, Slovakia
}

\date{\today}
%
%
%%%%%%%%%%%%%%%%%%%%%%%%%
\begin{abstract}
%%%%%%%%%%%%%%%%%%%%%%%%%
%
%
\vspace*{5mm}
We study for the first time the production of charmonia
in nuclear ultra-peripheral and electron-ion collisions 
based on a rigorous Green function formalism. 
Such formalism allows to incorporate properly 
formation effects (color transparency), as well as the quantum 
coherence inherent in higher twist shadowing corrections 
related to the $|Q\bar Q\ra$ Fock component of the photon. 
The leading twist gluon shadowing 
%{effect} 
associated with multi-gluon photon fluctuations 
{is} also included within the same formalism.
The later  
{effect} represents the dominant source of shadowing 
at mid rapidities in the LHC kinematic region, while the reduced  
{effect} of quark shadowing leads to a significant modification of differential 
cross sections $d\sigma/dy$ at forward and/or backwards rapidities.
Model calculations for $d\sigma/dy$ are in a good agreement with available 
UPC data on coherent charmonium production at RHIC and the LHC. 
In addition, we also perform predictions for nuclear effects 
in the electroproduction of charmonia, which can be verified 
by new data from electron-ion colliders.
%
%
%%%%%%%%%%%%%%%%%%
\end{abstract}
%%%%%%%%%%%%%%%%%%
%
%

%%%%%%%%%%%%%%%%%%%%%%%%%%%%%%%%%%
%%%%%%%%%%%%%%%%%%%%%%%%%%%%%%%%%%

\maketitle

%
%
%
%=============================================================
\section{Introduction}
\label{Sec:intro}
%=============================================================
%
%
%

Charmonium production in photo-nuclear reactions (photon virtuality $Q^2\approx 0$) 
is recently intensively studied in ultra-peripheral nuclear $Pb-Pb$ collisions (UPC) 
at the Large Hadron Collider (LHC), as well as in $Au-Au$ UPC collisions 
at Relativistic Heavy Ion Collider (RHIC).
{Moreover}, future experiments at planned Electron-Ion Collider (EIC) 
%----------------------------------------
\cite{AbdulKhalek:2021gbh,ATHENA:2022hxb}
%----------------------------------------
at RHIC will provide us with new data on coherent (elastic), $\gamma^*A\to V A$,
and incoherent (quasi-elastic), $\gamma^*A\to V A^*$ 
($A^*$ is an excited state of the A-nucleon system) charmonium 
($V = \Jpsi(1S)$, $\psip(2S)$,...), electroproduction ($Q^2\ne 0$).
This {will} allow {us} to 
{explore} manifestations of various nuclear phenomena, 
{such as} reduced effects of quantum coherence, color transparency, 
gluon shadowing, {and} gluon saturation {in new kinematic regions}.

The corresponding theoretical studies of nuclear effects are very effective 
within the light-front (LF) quantum-chromodynamics (QCD) color dipole approach
(see {e.g.,} our previous works 
%--------------------------------------------------------------------------------------------------
\cite{Kopeliovich:1991pu,Kopeliovich:1993gk,Kopeliovich:1993pw,Nemchik:1994fq,Nemchik:1996pp,Nemchik:1996cw,Kopeliovich:2001xj,Nemchik:2002ug,Ivanov:2002kc,Kopeliovich:2007wx,Kopeliovich:2008ek,Kopeliovich:2022jwe,Kopeliovich:2020has}
).
%--------------------------------------------------------------------------------------------------
%
This formalism allows to minimize uncertainties in model predictions 
for quarkonium yields on a proton target as was analyzed in
%----------------------------------------------------
Refs.~\cite{Krelina:2018hmt,Cepila:2019skb,Krelina:2019egg}
%----------------------------------------------------
(a comprehensive review on quarkonia phenomenology can be found e.g., in
%-------------------------------------------
Refs.~\cite{ivanov-04,brambilla-04,brambilla-10}). 
%-------------------------------------------
%
Such a reduction of theoretical uncertainties is then indispensable 
for a proper interpretation of phenomena occurring in electro-nuclear reactions.

In our previous work \cite{Kopeliovich:2020has}, we performed a detailed analysis 
of various effects accompanying the heavy quarkonium (HQ) production in UPC. 
Here we adopted several realistic $Q$-$\bar Q$ interaction potentials in 
the $Q\bar Q$ rest frame for calculations of realistic quarkonium wave functions 
by solving the Schr\"odinger equation with a subsequent  Lorentz boost to high energy. 
We also adopted several models for the phenomenological dipole cross sections 
fitted to deep-inelastic-scattering (DIS) data. 
We demonstrated that two main nuclear effects considerably affect 
the nuclear cross sections. 
In the RHIC and LHC kinematic regions, at forward and/or backward rapidities,
the first of them leads to the reduction of the higher twist quark shadowing 
related to the lowest $|Q\bar Q\ra$ Fock state of the photon.
The second effect causes the leading twist gluon shadowing, 
related to higher photon components containing gluons, 
in the mid rapidity region at the LHC energy range.

However, in
%-------------------------
Ref.~\cite{Kopeliovich:2020has}
%-------------------------
the reduced effects of quantum coherence 
{were} calculated only approximately.
Here, in the high energy limit of long coherence length, we provided additional 
corrections via the effective {form} factors modifying 
the nuclear cross sections at small and moderate c.m. 
photon energies, $W\lsim 30\div 40\,\GeV$. 
Although we have applied a rigorous path integral technique, 
the corresponding Green function acquired only the analytical 
harmonic oscillatory form reducing thus complexity of related calculations. 
Consequently, we {had to} use several simplifications, 
{such as} quadratic form for the dipole cross section, 
the harmonic oscillatory form for the light-front $Q$-$\bar Q$ 
interaction potential, and constant nuclear density.
Such simplifications cannot be applied for a sufficiently accurate 
description of data, especially on exclusive electroproduction 
of 2S-radially-excited heavy quarkonia as was demonstrated in 
%--------------------------
Ref.~\cite{Cepila:2019skb}.
%--------------------------

In this paper, we revise our previous calculations 
%--------------------------
\cite{Kopeliovich:2020has}.
%-------------------------- 
%
We present for the first time the model predictions for charmonium 
photo-nuclear and electro-nuclear production within the full 
quantum-mechanical description based on a rigorous path integral technique
%-------------------------
\cite{Kopeliovich:2001xj}.
%-------------------------
%
This allows to obtain more accurate results for reduced effects of quantum coherence 
beyond simplifications mentioned above. 
Besides the higher twist quark shadowing, the leading twist gluon 
shadowing effects are also calculated within the same Green
function formalism
%----------------------------------------------------------------------------------------------
\cite{Kopeliovich:1999am,Ivanov:2002kc,Kopeliovich:2008ek,Krelina:2020ipn,Kopeliovich:2022jwe}.
%-----------------------------------------------------------------------------------------------
%
This grants a proper and consistent incorporation of both the quantum coherence 
and color transparency effects. 
The precise calculation of these phenomena has then a huge impact for a conclusive 
evidence of saturation effects in UPC 
%----------------------------
\cite{Kovchegov:2023bvy}.
%----------------------------

In the most of recent papers (see {e.g.,}
%-------------------------------------------
Refs.~\cite{Cepila:2017nef,Henkels:2023plt}),
%-------------------------------------------
the effect of quantum coherence related to the $Q\bar Q$ 
photon fluctuations is not included properly, as well as shadowing corrections 
for higher multi-gluon Fock states of the photon are often neglected 
or incorporated inaccurately. 
%
%%%%%%%%%%%%%%%%%%%%%%%%%%%%%%%%%%%%%%%%%%%%%%%%%%%%%%%%%%
%\textbf{
Recently, new calculations at next-to-leading order (NLO)
of exclusive heavy quarkonium production 
have appeared in the literature (see e.g., Ref.~\cite{Mantysaari:2022kdm}).
However,
they have been performed only for the proton target
and their future extension to nuclear targets is not trivial 
and may be very complicated for analysis
of QCD phenomena.
For example, the latest NLO calculations of charmonium production
in UPC \cite{Eskola:2022vaf} do not include accurately the
quantum coherence effects in the whole rapidity region.
%}
%%%%%%%%%%%%%%%%%%%%%%%%%%%%%%%%%%%%%%%%%%%%%%%%%%%%%%%%%%
%
This provides another motivation for a more precise and consistent 
incorporation of nuclear effects in model predictions within the 
Green function formalism and represents thus the main goal of the present paper.
Due to scarcity of data on incoherent charmonium production and complexity 
of the corresponding calculations within the path integral technique, 
we will focus mainly on the analyses of model predictions for 
the elastic production of $\Jpsi(1S)$ and $\psip(2S)$. 
The diffractive incoherent (quasielastic) production of heavy quarkonia off nuclei
will be studied elsewhere.

The paper is organized as follows. 
In Sec.~\ref{Sec:cross-sec} we present the basic expression for the rapidity 
distributions $d\sigma/dy$ of heavy quarkonium production in heavy ion UPC. 
The Sec.~\ref{gf} is devoted to rigorous Green function formalism, 
which is applied for description of the coherent process.
We verify that such a formalism
gives a correct expression in the high energy limit. 
At large photon energies we can obtain thus the standard eikonal expression 
for the coherent quarkonium production cross section.
In Sec.~\ref{potential} we analyze the shape of the $Q$-$\bar Q$ interaction 
potential in the LF frame entering the Schr\"odinger equation for the 
Green function. 
We propose a simple prescription how to obtain such LF potentials from various
realistic models for the $Q$-$\bar Q$ interaction in the $Q\bar Q$ rest frame.
The resulting full formulas for the nuclear real and virtual photoproduction amplitudes, 
including also the effects of Melosh spin transformation, can be found in Sec.~\ref{melosh}.
Since the leading twist shadowing corrections have been analyzed and discussed in our previous works
%----------------------------------------------------------------------------------------------
\cite{Kopeliovich:1999am,Ivanov:2002kc,Kopeliovich:2008ek,Krelina:2020ipn,Kopeliovich:2022jwe},
%-----------------------------------------------------------------------------------------------
in Sec.~\ref{gs} we mention only briefly about the gluon shadowing, 
which is included in our calculations modifying the nuclear {photo- and} electroproduction 
cross sections at large photon energies.
In Sec.~\ref{res}, our model calculations within the color dipole approach based 
on a rigorous path integral technique are compared with available UPC data on charmonium coherent 
photoproduction off nuclear targets from experiments at RHIC and the LHC. 
Finally, we also present model calculations of nuclear cross sections in coherent charmonium
electroproduction in the kinematic regions covered by experiments at prepared EIC.
Our results are summarized and discussed in Sec.~\ref{conclusions}.

%
%
%
%============================================================
\section{Quarkonium production cross section in nuclear
ultra-peripheral and electron-ion collisions}
\label{Sec:cross-sec}
%============================================================
%
%
%

The photo-nuclear reaction in a heavy-ion UPC is generated 
by the photon field of one of the colliding heavy nuclei 
with {a} large charge $Z$, 
{which gives} rise to strong electromagnetic fields.
Then the Weizs\"acker-Williams photons are responsible for almost real 
photoproduction of a vector meson (quarkonium) $V$ with the cross section
derived in the one-photon-exchange approximation and
expressed in the rest frame of the target nucleus $A$ 
%----------------------
\cite{Bertulani:2005ru}
%----------------------
as
%
%====================================================================
\BE
  k\frac{d\sigma}{dk} = \int\,d^2\tau \int\,d^2b \,\,
  n(k,\vec b-\vec\tau,y)\, 
  \frac{d^2
  \sigma_A(s,b)}{d^2b}
  ~~ + ~~
  \Bigl \{ y\rightarrow -y \Bigr \}
  \,.
%--------------
\label{cs-upc}
%--------------
\EE
%====================================================================
%
%
Here the rapidity variable 
$y = \ln \bigl [s / (M_V \sqrt{s_N}~) \bigr] \approx \ln\bigl [(2 k M_N + M_N^2) / (M_V \sqrt{s_N}~)\bigr ]$ 
with $M_N$ and $M_V$ being the nucleon and vector meson mass, respectively, 
$\sqrt{s_N}$ is the collision energy and $k$ is the photon energy 
related to the square of the photon-nucleon center-of-mass (c.m.) energy 
$W^2 = 2 M_N k + M_N^2 - Q^2\approx s - Q^2$,
where $s$ is the {lepton}-nucleon c.m. energy squared.
In Eq.~(\ref{cs-upc}), the variable $\vec\tau$ is the relative 
impact parameter of a nuclear collision and $\vec{b}$ is the impact 
parameter of the photon-nucleon collision relative to the center of one of the nuclei. 
{For a collision of identical nuclei with the nuclear radius $R_A$ 
in UPC, it holds that $\tau > 2 R_A$ 
%-----------------------
\cite{Bertulani:2005ru}
%-----------------------
.}

The projectile nucleus induces the photon flux represented by the variable
$n(k,\vec b)$ in Eq.~(\ref{cs-upc}) with the following form 
%
%=================================================================
\BE
  n(k,\vec b) = \frac{\aem Z^2 k^2}{\pi^2\gamma^2}
  \Biggl [
  K_1^2\left(\frac{bk}{\gamma}\right)
  +
  \frac{1}{\gamma^2} K_0^2\left(\frac{bk}{\gamma}\right)
  \Biggr ]
  \,,
%--------------
  \label{flux}
%--------------
\EE
%==================================================================
%
where $\aem= 1/137.036$ is the fine-structure constant, $K_{0,1}$ are 
the modified Bessel functions of the second kind and the Lorentz factor 
$\gamma = 2 \gamma_{col}^2 - 1$ with $\gamma_{col} = \sqrt{s_N}/2 M_N$.
In heavy ion UPC at the LHC due to very large $\gamma\gg 1$
and small photon virtuality, $-q^2 = Q^2 < 1/R_A^2$,
the second term in Eq.~(\ref{flux}) {can be safely neglected} since
it corresponds to the flux of {longitudinally polarized} photons.

In the case of the photoproduction and electroproduction of heavy quarkonia 
{at} EIC {energies}, only the corresponding nuclear cross section 
$d^2\sigma_A(s,b)/d^2 b$ {is calculated directly}, 
i.e., without the convolution with the photon flux given by Eq.~(\ref{cs-upc}). 
Then total cross sections for quarkonium production on nuclear targets 
are given as an integral over the nuclear impact parameter $b$
%
%============================================
\BE
\sigma_A(s,b) = \int d^2 b ~~\frac{d^2\sigma_A(s,b)}{d^2 b}\,. 
\label{sigtot}
\EE
%============================================
%

%
%
%=================================================
\subsection{Coherent quarkonium production in the Green function formalism}
\label{gf}
%=================================================
%
%

In this paper, we study the elastic production of charmonia on nuclear targets 
{within the LF dipole approach, which has been applied to charmonium photoproduction off nucleons
%---------------------------------------------------
\cite{Hufner:2000jb,Krelina:2018hmt,Cepila:2019skb}
%---------------------------------------------------
and nuclei 
%-----------------------------------
\cite{Ivanov:2002kc,Nemchik:2002ug} before.}
%----------------------------------- 
%
The corresponding formula for the production cross section
has the following form \cite{Kopeliovich:2001xj}
%
%----------------------------------------------------------
\BA
%%%%%%%%%%%%%%%%%%%%%
  \label{sigcoh}
%%%%%%%%%%%%%%%%%%%%
   \frac{d^2\sigma_{A}^{coh}(s,b)}{d^2b}
%  \sigma_A^{coh}(s,b) 
  &=& \left|{\cal M}^{coh}(s,b)\right|^2\,,
%  \\
%%%%%%%%%%%%%%%%%%%%  
%  \label{siginc}
%%%%%%%%%%%%%%%%%%%% 
% \frac{d^2\sigma_A^{inc}(s,b)}{d^2b}
%%  \sigma_A^{inc}(s,b) 
%  &=& \frac{1}{16 \pi B(s,Q^2)}\,
%                           \int\limits_{-\infty}^{\infty} dz\,\rho_A(b,z)\,
%                           \left|{\cal M}^{inc}(s,b,z)\right|^2\,,
\EA
%-------------------------------------------------------
where the amplitude ${\cal{M}}^{coh}(s,b)$ reads
%-------------------------------------------------------
 \BA
%%   \Bigl|\,
%   {{\cal M}}^{inc}(s,b,z)
%%   \Bigr| 
%   &=&
%%       \Bigl|
%       H_{1}(s,{b},z) - H_{2}(s,{b},z)
%%       \Bigr|
%%%%%%%%%%%%%%%%%%%%
%   \label{gf-inc}\\
%%%%%%%%%%%%%%%%%%%%
   {{\cal M}}^{coh}({s,b}) 
   &=&
        \int\limits_{-\infty}^{\infty}\,dz\,\rho_{A}({b},z)\,
         H_{1}(s,{b},z)\, .
%%%%%%%%%%%%%%%%%%%
\label{gf-coh}
%%%%%%%%%%%%%%%%%%%
 \EA
%------------------------------------------------------
%
Here $\rho_A(b,z)$ is the nuclear density {distribution}, 
for which we employ the realistic Wood-Saxon form with parameters taken from {Ref.}
%---------------------
\cite{DeJager:1987qc}.
%---------------------

In Eq.~(\ref{gf-coh}), the function $H_{1}(s,b,z)$ 
can be expressed within a rigorous path integral technique as
%--------------------------------------------------------------------
 \BA
H_1(s,b,z) &=&
\int_0^1 d\alpha
\int d^{2} r_{1}\,d^{2} r_{2}\,
\Psi^{*}_{V}(\vec r_{2},\alpha)\,
G_{Q\bar Q}(z^{\prime}\to\infty,\vec r_{2};z,\vec r_{1})\,
\sigma_{Q\bar Q}(r_{1},s)\,
\Psi_{\gamma(\gamma^*)}(\vec r_{1},
\alpha)\,,
%\Bigl|_{z^\prime\to\infty}
%%%%%%%%%%%%%
\label{f1}
%\\
%%%%%%%%%%%%%
%
%H_{2}(s,b,z) &=& \frac{1}{2}\,
%\int\limits_{-\infty}^{z} dz_{1}\,\rho_{A}(b,z_1)\,
%\int\limits_0^1 d\alpha\int d^2 r_1\,
%d^2 r_{2}\,d^2 r\,
%\Psi^*_V (\vec r_2,\alpha)\nonumber \\
%&\times&
%G_{Q\bar Q}(z^{\prime}\to\infty,\vec r_2;z,\vec r)\,
%\sigma_{Q\bar Q}(\vec r,s)\,
%G_{Q\bar Q}(z,\vec r;z_1,\vec r_1)\,
%\sigma_{Q\bar Q}(\vec r_1,s)\,
%\Psi_{\gamma(\gamma^*)}(\vec r_1,\alpha)\, ,
%%%%%%%%%%%%
%\label{f2}
%%%%%%%%%%%%
 \EA
%---------------------------------------------------------------------
where 
$\Psi_V(\vec r,\alpha)$ is the LF wave function for heavy quarkonium and
$\Psi_{\gamma(\gamma^*)}(\vec r,\alpha)$ is the LF distribution 
of the $Q\bar Q$ Fock component of the quasi-real (transversely polarized) or virtual photon.
Both wave functions are dependent on the transverse separation $\vec{r}$ of the $Q\bar Q$ 
fluctuation (dipole) and on the variable $\alpha = p_Q^+/p_{\gamma}^+$, representing 
the boost-invariant fraction of the photon momentum carried by a heavy quark (or antiquark).

The universal dipole cross section $\sqq(r,s)$ in Eq.~(\ref{f1})
describes the interaction of the 
$Q\bar Q$ dipole with the nucleon target. 
Such interaction depends also on the c.m. energy squared $s$,
which can be alternatively included also via variable $x = (M_V^2+Q^2)/s$.

In Eq.~(\ref{f1}), the function $H_1(s,b,z)$ corresponds to 
the case when the incident photon produces coherently (elastic scattering) 
the colorless $Q\bar Q$ pair {at the point $z$}, 
which then evolves propagating through the nucleus and forms 
the heavy quarkonium wave function at $z^{\,\prime}\to\infty$.
{The} evolution of an interacting $Q\bar Q$ pair between points with particular 
initial and final longitudinal coordinates $z$ and $z^\prime$, and initial and final separations
$\vec r_1$ and $\vec r_2$, is described by the corresponding Green function
$G_{Q\bar Q}(z^\prime,\vec r_2;z,\vec r_1)$.
This Green function satisfies
the two-dimensional Schr\"odinger equation
%--------------------------------------------
\cite{Kopeliovich:1999am,Kopeliovich:2001xj}
%---------------------------------------------
%
%--------------------------------------------------------
 \BE
i\frac{d}{dz_2}\,G_{Q\bar Q}(z_2,\vec r_2;z_1,\vec r_1)=
\left[\frac{\eta^{2} - \Delta_{r_{2}}}{2\,k\,\alpha\,(1-\alpha)}
+V_{Q\bar Q}(z_2,\vec r_2,\alpha)\right]
G_{Q\bar Q}(z_2,\vec r_2;z_1,\vec r_1)\ .
\label{schroedinger}
 \EE
%-------------------------------------------------------
%
Here the Laplacian $\Delta_{r_2}$ acts on the coordinate $r_2$ and the variable
$\eta^2 = m_Q^2 + \alpha (1-\alpha) Q^2$, where $m_Q$ is the heavy quark mass.

Whereas the imaginary part of the the LF potential $V_{Q\bar Q}(z_2,\vec r_2,\alpha)$ in Eq.~(\ref{schroedinger}) controls the attenuation of the $Q\bar Q$ pair in the medium,
the corresponding real part describes the interaction between the $Q$ and $\bar{Q}$.
The particular shapes of ${\mathcal Re} V_{Q\bar Q}(z_2,\vec r_2,\alpha)$ 
are supposed to provide correct LF quarkonium wave functions.

The first kinetic term on the r.h.s. of Eq.~(\ref{schroedinger})
is responsible for the phase shift of the propagating $Q\bar Q$ pair.
The corresponding phase factor, $\exp[\,i \int_{z_1}^{z_2}\, dz q_L(z)]$,
contains the relative longitudinal momentum transfer $q_L(z)$ defined as
%
%------------------------------------------
\BA
q_L(z)
=
\frac{M_{Q\bar Q}^2(z) + Q^2}{2\,k}
=
\frac{\eta^2 + q_T^2}{2\,k\,\alpha(1-\alpha)}\,,
\EA
%------------------------------------------
%
where $q_T^2$ is the quark transverse momentum squared
and the effective mass squared of the $Q\bar Q$ pair reads
%
%-------------------------------------------------------------------
\BA
M_{Q\bar Q}^2 = \frac{m_Q^2 + q_T^2}{\alpha (1-\alpha)}\,.
\EA
%-------------------------------------------------------------------
%
Consequently, the {\it coherence time} (coherence length (CL))
$l_c = 1/q_L$ is contained in the kinetic term 
of the evolution equation (\ref{schroedinger}),
taking into account a replacement $q_T^2\Longrightarrow -\Delta_r$. 
Thus the CL is controlled by dynamically varying  
$Q\bar Q$ effective mass in Eq.~\eqref{schroedinger}.
Its static part can be seen in the last phase shift factor in the
Green function (\ref{gf-ho-re}).

For the HQ production, one can use a non-relativistic limit, $\alpha\sim 0.5$, $M_V^2\sim 4 m_Q^2$ and $m_Q^2\gg q_T^2$, leading to the following simplified form of the CL
%
%-----------------------------------------------------------------
\BA
\label{lc-hq}
l_c = \frac{2 k}{M_V^2 + Q^2}\,.
\EA
%-----------------------------------------------------------------
%
Such a length scale controls the effect of quantum coherence related to
the initial state higher twist shadowing, which causes suppression
as a result of destructive interference of amplitudes associated with interactions
on different bound nucleons.

Another phenomenon, accompanying the diffractive {photo- and} electroproduction of vector mesons 
(quarkonia) off nuclei, represents the final state absorption of produced quarkonia. 
It {is known as} the {\it color transparency} (CT).
Within the color dipole approach
(see {e.g.,}
%------------------------------------------------------------------------------------------------------------
Refs.~\cite{Kopeliovich:1991pu,Kopeliovich:1993gk,Kopeliovich:1993pw,Nemchik:1994fp,Nemchik:1994fq,Nemchik:1996cw,Hufner:2000jb,jan-00a,jan-00b}
%-------------------------------------------------------------------------------------------------------------
), the onset of CT is controlled by the $Q\bar Q$-size evolution 
during its propagation through the medium.
The 
%\sout{estimation} 
{estimate} of the corresponding time scale controlling such an evolution 
can be obtained in the rest frame of the nucleus from the uncertainty principle and reads 
%--------------------------------------------
\cite{Kopeliovich:1991pu,Kopeliovich:2001xj}
%--------------------------------------------
%
%-------------------------------------------
 \BE
t_f = \frac{2\,k}
{M_{V^\prime}^2 - M_V^2}\, ,
\label{tf}
 \EE
%--------------------------------------------
%
where $M_{V^\prime}$ is the mass of radially excited {quarkonium}.
The time scale $t_f$ in Eq.~(\ref{tf}) is usually called the 
{\it formation time} (formation length) and is also incorporated 
naturally in a rigorous quantum-mechanical description
presented in this work.

In our calculations, we include a small correction 
{in Eq.~(\ref{sigcoh})} due to the real part 
of the $\gamma^*~N\to V~N$ amplitude  
applying the following replacement 
%---------------------------------------------------
\cite{Bronzan:1974jh,Nemchik:1996cw,Forshaw:2003ki},
%---------------------------------------------------
%
%=======================================================
\BA
\sqq(r,s)
\Rightarrow
\sqq(r,s)
\,
\left(1 - i\,\frac{\pi}{2}\,
\Lambda
%\frac
%{\partial
 %\,\ln\,{\sqq(r,s)}}
%{\partial\,\ln s} 
\right),
\qquad\qquad
\Lambda = 
\frac
{\partial
 \,\ln\,{\sqq(r,s)}}
{\partial\,\ln s}
\, .
%--------------
  \label{re/im}
%--------------
\EA
%========================================================
%

Note that the amplitude (\ref{gf-coh}) is calculated 
within the Green function formalism without any restriction to CL. 
{In the high energy limit, when the CL $l_c\gg R_A$ ({\sl long coherence length} (LCL) regime)}, 
the Green function {acquires} a simple form, 
%
%--------------------------------------------------
 \BE
G_{Q\bar Q}(z_2,\vec r_2;z_1,\vec r_1) \Rightarrow
\delta^{(2)}(\vec r_1-\vec r_2)\,\exp\left[
- \frac{1}{2}\,\sigma_{Q\bar Q}(r_1,s)
\int\limits_{z_1}^{z_2} dz\,\rho_A(b,z)\right]\, .
\label{gf-lcl}
 \EE
%--------------------------------------------------
%
This is {the} so-called ``frozen'' approximation, since {the}
transverse sizes of such long-lived  $Q\bar Q$ photon fluctuations 
are frozen during propagation through the medium due to {the} Lorentz time dilation. 
Consequently, the LCL regime leads to the following simple form of the production amplitude (\ref{gf-coh}),
%
%-----------------------------------------------------------------
 \BE
{{\cal M}}^{coh}(s,b)
\Bigr|_{l_c \gg R_A} 
=
\int d^2r\int_0^1 d\alpha\,
\Psi_{V}^{*}(\vec r,\alpha)\,
\Biggl (1 -
\exp\left[-\frac{1}{2}\,\sqq(r,s)\,T_A(b)\right]
\Biggr )
\Psi_{\gamma(\gamma^*)}(\vec r,\alpha)\, ,
\label{coh-lcl}
 \EE
 %-----------------------------------------------------------------
 %
where $T_A(b) = \int_{-\infty}^{\infty} dz~\rho_A(b,z)$ represents the nuclear thickness function
normalized as $\int_{0}^{\infty} d^2 b~T_A(b) = A$, where $A$ is the nuclear mass number.

%
%
%============================================================
\subsection{The $Q$-$\bar Q$ potential {in the light-front frame}}
\label{potential}
%============================================================
%
%

Let's {consider} the propagation of an interacting $Q\bar Q$ pair in {a} vacuum, 
{where} the LF potential $V_{Q\bar Q}(z_2,\vec r_2,\alpha)$ in Eq.~(\ref{schroedinger}) 
contains only the real part.
The shape of ${\mathcal Re}V_{Q\bar Q}$ is strongly correlated with the form 
of the quarkonium wave function in the LF frame as discussed below.

We start from {the} determination of the well defined
quarkonium wave functions in the $Q\bar Q$ RF 
solving the Schr\"odinger equation with several realistic
potentials, such as the standard harmonic oscillatory potential (HO) 
(see {e.g.,}
%------------------------------
Ref.~\cite{Kopeliovich:1991pu} 
%------------------------------
), Cornell potential (COR) 
%------------------------------------
\cite{Eichten:1978tg,Eichten:1979ms},
%------------------------------------
logarithmic potential (LOG) 
%-------------------
\cite{Quigg:1977dd},
%-------------------
power-like potential (POW) 
%---------------------------------
\cite{Martin:1980jx,Barik:1980ai},
%---------------------------------
as well as Buchm\"uller-Tye potential (BT) 
%-------------------------
\cite{Buchmuller:1980su}. 
%-------------------------
%
All of them give rather close results for the charmonium electroproduction
cross sections as is presented in 
%-------------------------
Ref.~\cite{Hufner:2000jb} 
%-------------------------
and the corresponding differences in predictions can be treated as a measure of
theoretical uncertainty related to the LF quarkonium distribution function.
However, in our calculations we adopt only two of them, denoted as BT and POW, since 
they provide the best description of available data on proton targets, as was shown in 
%---------------------------
Ref.~\cite{Cepila:2019skb}.
%---------------------------

As the next step, the non-relativistic quarkonium wave functions in the $Q\bar Q$ RF 
are boosted to the LF frame using the standard Terent'ev prescription 
%----------------------
\cite{Terentev:1976jk} 
%----------------------
as described in 
%-------------------------------------------------------------------------
Refs.~\cite{Hufner:2000jb,Krelina:2018hmt,Cepila:2019skb,Krelina:2019egg} 
%-------------------------------------------------------------------------
(see also Sec. II.B in 
%---------------------------------
Ref.~\cite{Kopeliovich:2022jwe}).
%---------------------------------
%
Such a prescription is based on {the} unjustified assumption that, 
after performing the Lorentz boost to the LF frame,
{where} the original quarkonium wave function in the RF is expressed 
in terms of light-front and Lorentz invariant variables 
related to 2-dimensional transverse momentum $\vec q_T$ and $\alpha$, 
the {resulting} wave function is also Lorentz invariant.
{Nevertheless}, this Lorentz boosting prescription 
{was} tested and verified in
%------------------------------
Ref.~\cite{Kopeliovich:2015qna}
%------------------------------
using the Lorentz boosted Schr\" odinger equation for heavy $Q\bar Q$ 
system based on the Lorentz-invariant Bethe-Salpeter equation.

As soon as the ground-state quarkonium wave function is known in the LF frame, 
one can subsequently determine the corresponding real part of the LF potential 
$V_{Q\bar Q}(z,\vec r,\alpha)$. 
Let's consider for simplicity the harmonic oscillatory potential
in the $Q\bar Q$ RF,
%
%------------------------------------------------------------
\BE
V_{Q\bar Q}^{HO}(\rho) = \frac{1}{2}\,m_Q\,\omega_0^2\,\rho^2\, ,
\label{pot-osc}
\EE
%------------------------------------------------------------
%
where $\rho$ is the 3-dimensional separation between $Q$ 
and $\bar Q$ and $\omega_0\approx 0.3\,\GeV$ is the oscillatory frequency. 
The Schr\"odinger equation with this potential has an analytical solution 
leading to the following Gaussian shape of the quarkonium wave function in the RF
%
%------------------------------------------------------------
\BE
\Psi_V^{HO}(\rho) = N_{HO}\cdot \exp\Bigl [ - \frac{1}{4}\,m_Q\,\omega_0\,\rho^2\Bigr ]\,,
\EE
%-------------------------------------------------------------
%
where $N_{HO}$ is the normalization factor such 
{that} $\int d^3 \rho~|\Psi^{HO}(\rho)|^2 = 1$.

Applying the recipe from 
%----------------------------
Ref.~\cite{Terentev:1976jk}, 
%----------------------------
as a result of the transition from the nonrelativistic $Q\bar Q$ rest frame to the LF frame, one can obtain the following LF counterpart 
%---------------------------------------
\cite{Nemchik:1996cw,Kopeliovich:2001xj}
%---------------------------------------
of $\Psi_V^{HO}(\rho)$
%
%----------------------------------------------------------------
\BE
\Psi_V^{HO}(\vec r,\alpha) = 
C\,\alpha (1-\alpha)\,f(\alpha)\,\exp\Bigl [- \frac{1}{2} a^2(\alpha) r^2\Bigr ]
\label{psi-lc}
\EE
%-----------------------------------------------------------------
%
with
%
%-------------------------------------------------------------------
\BE
f(\alpha) = \exp\Bigl [ - \frac{m_Q^2}{2 a^2(\alpha)}
                        + \frac{4 \alpha (1-\alpha) m_Q^2}{2 a^2(\alpha)} \Bigr ]\, ,
\EE
%-------------------------------------------------------------------
%
and
%
%-------------------------------------------------------------------
\BA
a^2(\alpha) = \omega_0\,m_q\,4\alpha(1-\alpha)/2\ = a_1^2 4 \alpha(1 - \alpha),
\EA
%-------------------------------------------------------------------
%
keeping the normalization condition $\int d^2 r\,d \alpha ~|\Psi_V^{HO}(\vec r,\alpha)|^2 = 1$.

In order to find the real part of the LF potential $V_{Q\bar Q}^{HO}(z,\vec r,\alpha)$ 
in the LF frame, the wave function (\ref{psi-lc}) should satisfy the following Schr\"odinger equation 
%
%--------------------------------------------------------------------
\BA
    \left( \frac{a^2(\alpha)-\Delta_r}{2\,k\,\alpha(1-\alpha)} + {\cal R}e V_{Q\bar Q}^{HO}(z,\vec r,\alpha) \right) 
    \Psi^{HO}_V(\vec r,\alpha) = E_{LF}^{HO}\cdot \Psi^{HO}_V(\vec r,\alpha)\,,
\label{schroedinger-ho}
\EA
%--------------------------------------------------------------------
%
with the corresponding eigenvalue $E_{LF}^{HO} = 3\,a^2(\alpha) / \bigl (2\,k\,\alpha(1-\alpha)\bigr)$. 
This equation can be solved analytically giving the following form of ${\mathcal Re} V_{Q\bar Q}(z,\vec r,\alpha)$ \cite{Kopeliovich:1999am}
%
%--------------------------------------------------------------------
\BE
{\mathcal Re}\,V_{Q\bar Q}(z,\vec r,\alpha) =
\frac{a^4(\alpha)\,r^2}
{2\,k\,\alpha(1-\alpha)}\ .
\label{pot-real}
 \EE
%-----------------------------------------------------------------------
%
Indeed, such a real part of the LF potential was used in 
%----------------------------------------------------------------
Refs.~\cite{Kopeliovich:2001xj,Nemchik:2002ug,Kopeliovich:2007wx}
%----------------------------------------------------------------
for {the} study of shadowing and absorption phenomena in vector meson electroproduction 
off nuclear targets within a rigorous Green function formalism.
The shape of the LF wave function {in} Eq.~(\ref{psi-lc}) guarantees a correct 
non-relativistic limit ($\alpha \to 0.5$, $\vec r\to\vec\rho$) of the LF potential, 
Eq.~(\ref{pot-real}), leading to the HO form in the RF given by Eq.~(\ref{pot-osc}).

Considering now an arbitrary model for $Q$-$\bar Q$ interaction potential in the $Q\bar Q$ RF,
one can obtain ${\mathcal Re}\,V_{Q\bar Q}(z,\vec r,\alpha)$ in the LF frame only numerically 
as a solution of the following Schr\"odinger equation, which is similar to Eq.~(\ref{schroedinger-ho}),
%
%-----------------------------------------------------
\BA
    \left( \frac{A^2(\alpha)-\Delta_r}{2\,k\,\alpha(1-\alpha)} + 
    {\mathcal Re} V_{Q\bar Q}(z,\vec r,\alpha) \right) \Psi_V(\vec r,\alpha) = E_{LF}\cdot \Psi_V(\vec r,\alpha)\,,
\label{eq:SchEqLC}
\EA
%-----------------------------------------------------
%
where the LF potential including the term with function $A^2(\alpha)$ can be written as
%
%---------------------------------------------
\BA
 {\mathcal Re} V^*_{Q\bar Q}(z,\vec r,\alpha) 
 =   
 {\mathcal Re} V_{Q\bar Q}(z,\vec r,\alpha) 
 + 
 \frac{A^2(\alpha)}{2\,k\,\alpha(1-\alpha)} 
 = 
 E_{LF}
 +
 \frac{1}{\Psi_V(\vec r,\alpha)}\cdot\frac{\Delta_r\, \Psi_V(\vec r,\alpha)}{2\,k\,\alpha(1-\alpha)}~.
\EA
%---------------------------------------------
%
In analogy with the HO potential, the function $A^2(\alpha) = A_1^2\,4\,\alpha(1-\alpha)$ 
and the factor $A_1^2$ can be determined from the non-relativistic limit 
(
{$\vec r\to\vec \rho, \alpha\to 0.5$}) of Eq.~\eqref{eq:SchEqLC},
%
%-------------------------------------------------------------
\BA
    A_1^2
    = 
    m_Q\,\bigl [E_{RF} - V_{Q\bar{Q}}(\vec \rho)\bigr ] + \frac{\Delta_r\, \Psi_V(\vec r\to\vec \rho,\alpha\to 0.5)}{\Psi_V(\vec r\to\vec\rho,\alpha\to 0.5)}\,,
\EA%----------------------------------------------------------
where $V_{Q\bar Q}(\vec{\rho})$ is the quark interaction potential 
in the $Q\bar Q$ rest frame and the eigenvalue $E_{RF}$ 
is related to that in the LF frame via relation,
%
%----------------------------------
\BA
    E_{LF} = E_{RF}\cdot\frac{m_Q\,4\,\alpha(1-\alpha)}{2\,k\,\alpha(1-\alpha)}\,.
\EA
%---------------------------------
%
Finally, the LF potential for any model describing the $Q$-$\bar{Q}$ interaction 
in the RF can be obtained from the following relation
%
%-------------------------------------------------
\BA
\label{re-pot}
{\mathcal Re} V_{Q\bar Q}(\vec r,\alpha) = E_{LF}+\frac{1}{\Psi_V(\vec r,\alpha)}\cdot\frac{\Delta_r \Psi_V(\vec r,\alpha)}{2\,k\,\alpha(1-\alpha)} - \frac{A_1^2\,4\,\alpha(1-\alpha)}{2\,k\,\alpha(1-\alpha)}\,.
\EA
%-------------------------------------------------
%

Taking the analytical form of ${\mathcal Re}\,V_{Q\bar Q}(z_2,\vec r_{2},\alpha)$  given by Eq.~(\ref{pot-real}) corresponding to the harmonic oscillatory form (\ref{pot-osc}) in the $Q\bar Q$ rest frame, the evolution equation (\ref{schroedinger}) has an analytical solution, the harmonic oscillator Green function \cite{fg}
%
%---------------------------------------------------------------------
 \BA
G_{Q\bar Q}(z_2,\vec r_2;z_1,\vec r_1) 
&=&
\frac{a^2(\alpha)}{2\;\pi\;i\;
{\rm sin}(\omega\,\Delta z)}\, {\rm exp}
\left\{\frac{i\,a^2(\alpha)}{2~{\rm sin}(\omega\,\Delta z)}\,
\Bigl[(r_1^2+r_2^2)\,{\rm cos}(\omega \;\Delta z) -
2\;\vec r_1\cdot\vec r_2\Bigr]\right\}
\nonumber\\ 
&\times&
{\rm exp}\left[-
\frac{i\,\eta^{2}\,\Delta z}
{2\,k\,\alpha\,(1-\alpha)}\right] \ ,
\label{gf-ho-re}
 \EA
%---------------------------------------------------------------------
%
where $\Delta z=z_2-z_1$ and
%
%---------------------------------------------------------------------
 \BE \omega = \frac{a^2(\alpha)}{k\;\alpha(1-\alpha)}\ .
\label{omega-lc}
 \EE
%---------------------------------------------------------------------

The boundary condition in Eq.~(\ref{gf-ho-re}) is $G_{Q\bar Q}(z_2,\vec r_2;z_1,\vec r_1)|_{z_2=z_1}=
\delta^{(2)}(\vec r_1-\vec r_2)$. 
For other LF potentials ${\mathcal Re}\,V_{Q\bar Q}(z_2,\vec r_{2},\alpha)$, 
corresponding to COR, LOG, POW and BT models in the $Q\bar Q$ rest frame,
Eq.~(\ref{schroedinger}) for the Green function {has} to be solved only numerically.
Note that the LF oscillatory frequency $\omega$ in Eq.~(\ref{omega-lc}) has 
a correct non-relativistic limit $\omega_0$ in the $Q\bar Q$ rest frame.

Considering the propagation of an interacting $Q\bar Q$ pair in the nuclear medium,
the corresponding Green function satisfies the same evolution equation as given by
Eq.~(\ref{schroedinger}).
However, the LF potential acquires besides the real part 
(see e.g., Eq.~(\ref{pot-real}) for the HO form of the LF potential
) 
also the imaginary part, which represents an absorption of the $Q\bar Q$ 
Fock state of the photon in the medium,
%
%-------------------------------------------------------------------
 \BE
{\mathcal Im} V_{Q\bar Q}(z_2,\vec r,\alpha) = -
\frac{\sigma_{Q\bar Q}(\vec r,s)}{2}\,\rho_{A}({b},z_2)\,.
\label{pot-im}
 \EE 
%-------------------------------------------------------------------
%
Such form of the LF potential was used in 
%---------------------------------------------------------------------------------------------------
Refs.~\cite{Kopeliovich:1998gv,Kopeliovich:2000ra,Nemchik:2003wx,Kopeliovich:2008ek,Krelina:2020ipn}
%---------------------------------------------------------------------------------------------------
for calculation of nuclear shadowing in deep-inelastic scattering off nuclei in a good agreement with data.

In order to keep the analytical solution of Eq.~(\ref{schroedinger}), the whole LF potential 
$V_{Q\bar Q}(z_2,\vec r_{2},\alpha)$ should acquire the HO form, 
i.e., the imaginary part given by Eq.~(\ref{pot-im}) should be proportional to $r^2$ {as well}. 
This requires to employ the quadratic $r$-dependence of the dipole cross section
%
%-------------------------------------------------------------
 \BE
\sigma_{Q\bar Q}(r,s) = C(s)\,r^2\, .
\label{dcs-r2}
 \EE
%-------------------------------------------------------------
%
Such an approximation has a reasonable good accuracy for heavy quarkonium 
photoproduction and electroproduction due to sufficiently small $Q$-$\bar Q$ dipole sizes 
contributing to production amplitudes. 
However, instead of the approximation (\ref{dcs-r2}), in the present paper we adopt 
two realistic phenomenological parametrizations for $\sqq(r,s)$ of the saturated form.
Model predictions with {the parametrization for}
$\sqq(r,s)$ from {Ref.} \cite{Golec-Biernat:2017lfv} within 
the known Golec-Biernat-Wushoff (GBW) model 
%-----------------------------------------------
\cite{GolecBiernat:1998js,Golec-Biernat:1999qor}
%-----------------------------------------------
will be compared with results based on Kopeliovich-Schafer-Tarasov (KST) model 
%-------------------------
\cite{Kopeliovich:1999am},
%-------------------------
extended for the additional Reggeon term as given in 
%------------------------------
Ref.~\cite{Kopeliovich:2001xj}.
%------------------------------

%%%%%%%%%%%%%%%%%%%%%%%%%%%%%%%%%%%%%%%%%%%%%%%%%%%%%%%%%%%%%%%%%%%%%%%%%%%%%%%%%%%%
       %%%%%%%%%%%%%%%%%%%%%%%%%%%%% FIG. A %%%%%%%%%%%%%%%%%%%%%%%%%%%%%%%%
%%%%%%%%%%%%%%%%%%%%%%%%%%%%%%%%%%%%%%%%%%%%%%%%%%%%%%%%%%%%%%%%%%%%%%%%%%%%%%%%%%%%
\BF
%%%%%%%%%%%%%%%%
\PSfig{0.35}{dsdyGF_psi1S_200Au_BT_KSTr_Re_ImV_newGS.eps} \hspace{40pt}
\PSfig{0.35}{dsdyGF_psi2S_200Au_BT_KSTr_Re_ImV_newGS.eps}
\\
\vspace*{-0.20cm}
\Caption{
%------------------
\label{Fig-UPCpsi-pot}
%------------------
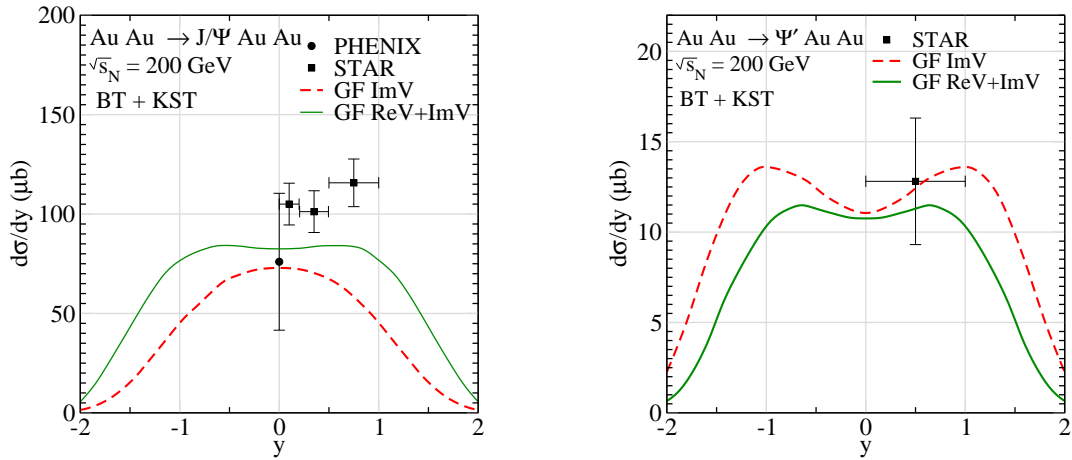
(Color online) 
     The impact of ${\mathcal Re} V_{Q\bar Q}$ on
     rapidity distribution $d\sigma/dy$ for coherent $\Jpsi(1S)$ (left panel)
     and $\psip(2S)$ (right panel) production in UPC at
     c.m. collision energy $\sqrt{s_N} = 200\,\GeV$.
     Values of $d\sigma/dy$ are calculated within the Green function formalism adopting
     the BT $Q$-$\bar Q$ interaction potential and KST model for dipole cross section.
     Dashed and solid lines correspond to the case when ${\mathcal Re} V_{Q\bar Q}$ is neglected
     and included in model predictions, respectively.
     The data are from the PHENIX \cite{Afanasiev:2009hy} and  
     STAR \cite{STAR:2023nos,STAR:2023gpk} experiments.
     }
\EF
%%%%%%%%%%%%%%%%%%%%%%%%%%%%%%%%%%%%%%%%%%%%%%%%%%%%%%%%%%%%%%%%%%%%%%%%%%%%%%%%%%%%

Note that the relevance of ${\mathcal Re} V_{Q\bar Q}$ in calculations of charmonium yields
gradually decreases with the photon energy as one can see from Eq.~(\ref{re-pot}).
Consequently, the modification of $d\sigma/dy$ related to ${\mathcal Re} V_{Q\bar Q}$ 
manifests itself only at large forward and/or backward rapidities and is more visible 
at smaller RHIC energies compared to {the} LHC kinematic region.
In Fig.~\ref{Fig-UPCpsi-pot}, we compare model predictions 
based on the path integral technique at c.m. collision energy 
$\sqrt{s_N} = 200\,\GeV$ neglecting (dashed lines) and including (solid lines) 
${\mathcal Re} V_{Q\bar Q}$.
One can see that the importance of ${\mathcal Re} V_{Q\bar Q}$ gradually
decreases towards the mid rapidity region $y=0$.
Whereas for the coherent $\Jpsi(1S)$ production in UPC the additional
incorporation of ${\mathcal Re} V_{Q\bar Q}$ enhances $d\sigma/dy$, 
the nodal structure of the $\psip(2S)$ wave function leads
to a counter-intuitive reduction of $d\sigma/dy$.

%
%
%==================================================================
\subsection{Nuclear production amplitudes incorporating Melosh spin effects}
\label{melosh}
%==================================================================
%
%

Terent'ev boosting prescription 
%---------------------
\cite{Terentev:1976jk}
%---------------------
to the LF frame requires a significant correction 
%--------------------
\cite{Hufner:2000jb} 
%--------------------
due to transverse motion of the quarks when their
momenta are not parallel to the boost axis.
This leads to quark spin rotation effects known also as {the} Melosh spin effects 
%--------------------
\cite{Melosh:1974cu} 
%--------------------
that have been included in our calculations following 
%---------------------------------------------------------------
Refs.~ \cite{Krelina:2018hmt,Lappi:2020ufv,Kopeliovich:2022jwe}.
%---------------------------------------------------------------
%
We also {ruled} out the photon-like vertex for a heavy 
quarkonium transition $V\to Q\bar Q$ pair due to abnormally 
large weight of a $D$-wave component, as analyzed in 
%--------------------------------
Ref.~\cite{Kopeliovich:2022jwe}.
%--------------------------------

In this paper, we have adopted the structure of the $V\to Q\bar Q$ vertex from our previous studies in 
%--------------------------------------------------------------------------------------------
Refs.~\cite{Ivanov:2002kc,Ivanov:2007ms,Krelina:2018hmt,Cepila:2019skb,Kopeliovich:2022jwe}.
%--------------------------------------------------------------------------------------------
%
Then, assuming $s$-channel helicity conservation,
the final expression for the coherent amplitude including the Melosh spin rotation is given 
{again} by Eq.~(\ref{gf-coh}) but with the function $H_{1}(s,b,z)$ of the following form
%
%-----------------------------------------------------
\BA
H_1^{T,L}(s,b,z) &=&
\int_0^1 d\alpha
\int d^{2} r_{1}\,d^{2} r_{2}\,
%\Psi^{*}_{V}(\vec r_{2},\alpha)\,
G_{Q\bar Q}(z^\prime,\vec r_{2};z,\vec r_{1})
\Bigl|_{z^\prime\to\infty}\,
%\nonumber \\
%&\times&
\cdot\,
%\Bigl [
\Omega_{T,L}(r_1,r_2,\alpha)\cdot
%\Bigr ]
\,
\sqq(r_{1},s)\, ,
%\Psi_{\gamma}(\vec r_{1},\alpha)
\label{f1-sr}
\EA
%-----------------------------------------------------
%
%
%-----------------------------------------------------
%\BA
%H_{2}^{T,L}(s,b,z) &=& 
%\frac{1}{2}\,
%\int\limits_{-\infty}^{z} dz_{1}\,\rho_{A}(b,z_1)\,
%\int\limits_0^1 d\alpha\int d^2 r_1\,
%d^2 r_{2}\,d^2 r\,
%%\Psi^*_V (\vec r_2,\alpha)\nonumber \\
%G_{Q\bar Q}(z^{\prime}\to\infty,\vec r_2;z,\vec r)\,
%\sqq(\vec r,s)\,
%\nonumber\\
%&\times&
%G_{Q\bar Q}(z,\vec r;z_1,\vec r_1)\,
%\cdot
%%\Bigl [
%%\Sigma^{(1)}(r_1,r_2,\alpha) + \Sigma^{(2)}(r_1,r_2,\alpha) 
%\Omega_{T,L}(r_1,r_2,\alpha)\cdot
%%\Bigr ]
%\,
%\sqq(\vec r_1,s)\,,
%%\Psi_{\gamma}(\vec r_1,\alpha)\, ,
%\label{f2-sr}
%\EA
%------------------------------------------------------
where
$\Omega_T(r_1,r_2,\alpha) = \Sigma_T^{(1)}(r_1,r_2,\alpha) + \Sigma_T^{(2)}(r_1,r_2,\alpha)$ 
and
$\Omega_L(r_1,r_2,\alpha) = \Sigma_L(r_1,r_2,\alpha)$
for transversely (T) and longitudinally (L) polarised photon and quarkonium, respectively.
Treating charmonium electroproduction, one can safely 
{consider}, with a rather good accuracy,
the perturbative wave functions for the incoming photon, 
$\Psi_{\gamma(\gamma^*)}(\vec r,\alpha) = K_0(\eta\,r)$. 
Then the functions $\Sigma_T^{(1,2)}$ and $\Sigma_L$ read
%
%=================================================================
\BA
  \Sigma^{(1,2)}_T(r_1,r_2,\alpha)
   &=&  
   N\,
   K_{0,1}(\eta\,r_1) \int\limits_{0}^{\infty} dp_T\,p_T\,
   J_{0,1}(p_T r_2) \Psi_V (p_T,\alpha) \,
    \mathcal{R}_T^{(1,2)}(p_T)
%-------------------
\label{sigma-sr1}
%-------------------
\EA
%==================================================================
%
with
%
%===============================================================
\BA
\mathcal{R}_T^{(1)}(p_T) 
%&=& 
=
\frac{2\,m_Q^2(m_L+m_T)+m_L\,p_T^2}{ m_T (m_L + m_T)} \,,
\qquad\qquad\quad
\mathcal{R}_T^{(2)}(p_T) 
&=&
\frac{m_Q^2(m_L+2m_T)-m_T\,m_L^2}{m_Q^2\,m_T (m_L+m_T)}\, \eta\,p_T\,,
%------------------
\label{sigma-sr2}
%------------------
\EA
%==============================================================
%
and
%
%=================================================================
\BA
  \Sigma_L(r_1,r_2,\alpha)
   &=&  
   N\,
   K_{0}(\eta\,r_1) \int\limits_{0}^{\infty} dp_T\,p_T\,
   J_{0}(p_T r_2) \Psi_V (p_T,\alpha) \,
    \mathcal{R}_L(p_T)
%-------------------
\label{sigmaL-sr1}
%-------------------
\EA
%==================================================================
%
with
%
%===============================================================
\BA
\mathcal{R}_L(p_T) 
%&=& 
=
4 Q \alpha (1-\alpha)\,
\frac{m_Q^2 + m_L m_T}{m_Q (m_L + m_T)} \,.
%------------------
\label{sigmaL-sr2}
%------------------
\EA
%==============================================================
%
Here $N=Z_Q\,\sqrt{2 N_c \,\alpha_{em}}/4\,\pi$, 
where the factor $N_c=3$ represents the number of colors in QCD, 
$Z_Q=\frac{2}{3}$ 
is the charge-isospin factor for the production of charmonia,  
and $J_{0,1}$ are the Bessel functions of the first kind.
The variables $m_{T,L}$ in Eqs.~(\ref{sigma-sr2}) and (\ref{sigmaL-sr2})
have the following form
%
%------------------------------------------------------
\BE
m_T = \sqrt{m_Q^2 + p_T^2} \,, 
\qquad m_L = 2\, m_Q\,\sqrt{\alpha(1-\alpha)}\,.
\EE
%------------------------------------------------------
%
The magnitude of the charm quark mass corresponds to values from the realistic 
phenomenological models for the $c$-$\bar c$ interaction potential, 
such as POW and BT, used in our analysis.

Finally, the total cross section (\ref{sigtot}) can be expressed as the sum of T and L contributions
%
%*************************************************
\BA
\sigma_A^{coh}(s) 
=
\sigma_{A,T}^{coh}(s)
+
\tilde{\varepsilon}\,
\sigma_{A,L}^{coh}(s)
\,,
%%%%%%%%%%%%%%%%
\label{total-cs}
%%%%%%%%%%%%%%%%
\EA
%************************************************
%
where we have taken the photon polarization $\tilde{\varepsilon} = 1$.

%
%
%
%=======================================
\subsection{Gluon shadowing}
\label{gs}
%=======================================
%
%
%

It was shown in 
%----------------------------------------------
Refs.~\cite{Ivanov:2007ms, Kopeliovich:2020has} 
%----------------------------------------------
that heavy quarkonia produced in photo-nuclear reactions in UPC 
are affected mainly by two specific nuclear effects: 
the gluon shadowing (GS) and the effect of reduced coherence length when $l_c\lsim R_A$. 
Whereas the later effect is naturally incorporated in the Green function 
formalism described  
in Sec.~\ref{gf}, the effects of GS 
{have} to be included additionally as a shadowing correction corresponding to higher Fock
components of the photon containing gluons, 
i.e., $|Q\bar QG\ra$, $|Q\bar Q2G\ra$ ... $|Q\bar QnG\ra$.

The leading twist gluon shadowing was introduced in 
%------------------------------
Ref.~\cite{Kopeliovich:1999am} 
%------------------------------
within the dipole representation.
We have demonstrated in 
%-------------------------------------------------------------------------------
Refs.~\cite{Kopeliovich:2001xj,Ivanov:2002kc,Nemchik:2002ug,Kopeliovich:2007wx}
%-------------------------------------------------------------------------------
that such effect substantially modifies cross sections at high energies
in photoproduction of vector mesons on nuclei.
According to analysis and discussion in 
%-------------------------------
Ref.~\cite{Kopeliovich:2022jwe},
%-------------------------------
only the one-gluon Fock state $|Q\bar QG\ra$ of the photon gives a dominant
contribution to {the} nuclear shadowing within the kinematic regions 
of present UPC experiments at the LHC.

We have incorporated the GS correction in our calculations as 
a reduction of $\sqq(r,s)$ in nuclear reactions with respect 
to processes on the nucleon
%-------------------------
\cite{Kopeliovich:2001ee},
%-------------------------
%
%------------------------------------------------------------
\BE
  \sqq(r,x) \Rightarrow \sqq(r,x) \cdot R_G(x,b)\,,
\label{eq:dipole:gs:replace}
\EE
%------------------------------------------------------------
%
where
the correction factor $R_G(x,b)$ 
related to the $Q\bar QG$ component of the photon, was calculated 
within the Green function formalism 
{as a function of the nuclear impact parameter $b$ and variable $x$,} 
{in} analogy with our previous works
%----------------------------------------------------------------------------------------------------------
\cite{Kopeliovich:1999am,Kopeliovich:2001xj,Kopeliovich:2001ee,Ivanov:2002kc,Nemchik:2002ug,Kopeliovich:2008ek,Krelina:2020ipn}
%----------------------------------------------------------------------------------------------------------
(see also Fig.~1 and discussion in 
%-------------------------------
Ref.~\cite{Kopeliovich:2022jwe}).
%-------------------------------

%
%
%
%====================================================
\section{Model predictions vs available data}
\label{res}
%====================================================
%
%
%

We calculated the coherent charmonium photoproduction in UPC according 
to Eq.~(\ref{cs-upc}), as well as electroproduction {at} EIC {energies} 
using Eqs.~(\ref{sigtot}) and (\ref{total-cs}).
The corresponding nuclear cross sections are obtained within 
a rigorous Green function formalism as described in Sec.~\ref{gf}. 
The evolution equation, Eq.~(\ref{schroedinger}), for the Green function 
is solved numerically following the algorithm from 
%---------------------------
Ref.~\cite{Nemchik:2003wx}.
%---------------------------
%
The Melosh spin transformation is included as specified in Sec.~\ref{melosh}. 
Besides effects of the quantum coherence for the lowest $|Q\bar Q\ra$ Fock state 
of the photon, we included {also the leading twist gluon shadowing 
corrections (see Sec.~\ref{gs})} in our {calculations} at large photon energies.

%%%%%%%%%%%%%%%%%%%%%%%%%%%%%%%%%%%%%%%%%%%%%%%%%%%%%%%%%%%%%%%%%%%%%%%%%%%%%%%%%%%%
       %%%%%%%%%%%%%%%%%%%%%%%%%%%%% FIG. 1 %%%%%%%%%%%%%%%%%%%%%%%%%%%%%%%%
%%%%%%%%%%%%%%%%%%%%%%%%%%%%%%%%%%%%%%%%%%%%%%%%%%%%%%%%%%%%%%%%%%%%%%%%%%%%%%%%%%%%
\BF
%%%%%%%%%%%%%%%%
\hspace*{-0.3cm}
\PSfig{0.36}{dsdyGF_psi1S_200Au_Re+ImV_newGS.eps}
\hspace*{1.50cm}
\PSfig{0.36}{dsdyGF_psi2S_200Au_Re+ImV_newGS.eps}
\\
\PSfig{0.36}{dsdyGF_psi1S_2760Pb_Re+ImV_newGS.eps}
\hspace*{1.50cm}
\PSfig{0.36}{dsdyGF_psi2S_2760Pb_Re+ImV_newGS.eps}
\\
\PSfig{0.36}{dsdyGF_psi1S_5020Pb_Re+ImV_newGS.eps}
\hspace*{1.50cm}
\PSfig{0.36}{dsdyGF_psi2S_5020Pb_Re+ImV_newGS.eps}
\\
\vspace*{-0.0cm}
\Caption{
%------------------
\label{Fig-UPC1psi}
%------------------
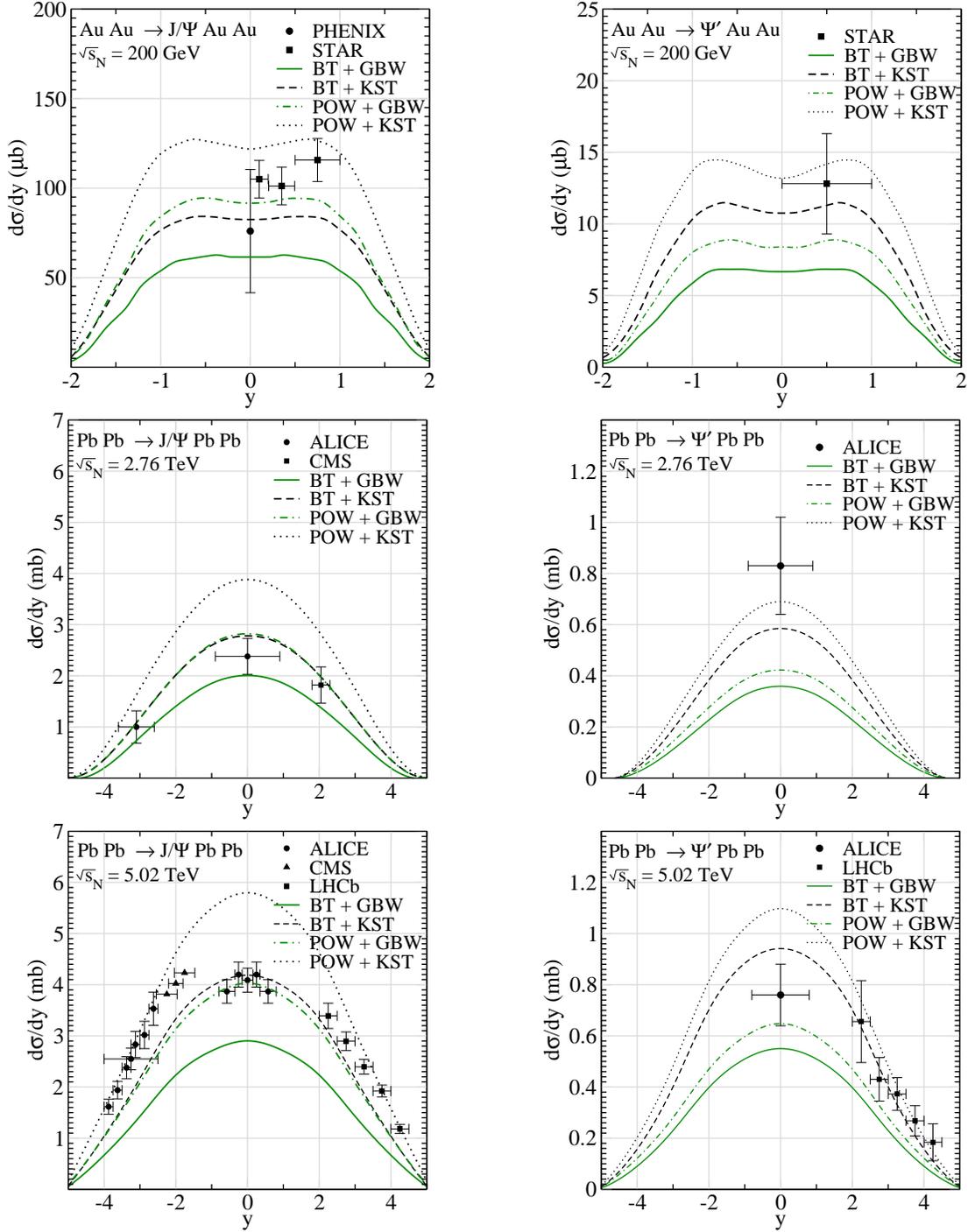
(Color online)
    Rapidity distributions of coherent $\Jpsi(1S)$ (left panels) and $\psip(2S)$ (right panels)
    photoproduction in UPC at RHIC collision energy $\sqrt{s_N}=200\,\GeV$ (top panels) and at 
    LHC energies $\sqrt{s_N} = 2.76\,\TeV$ (middle panels), and $\sqrt{s_N} = 5.02\,\TeV$ 
    (bottom panels). 
    The nuclear cross sections are calculated with charmonium wave functions 
    generated by the POW (dot-dashed and dotted lines) and BT (solid and dashed lines) potentials 
    {using} GBW (solid and dot-dashed lines) and KST (dashed and dotted lines) 
    models for the dipole cross section. 
    The data for $d\sigma/dy$ from the PHENIX \cite{Afanasiev:2009hy},  
    STAR \cite{STAR:2023nos,STAR:2023gpk}, CMS \cite{Khachatryan:2016qhq}, 
    ALICE \cite{Abelev:2012ba,Abbas:2013oua,Adam:2015sia,Acharya:2019vlb,ALICE:2021gpt},  
    LHCb \cite{LHCb:2018ofh,LHCb:2022ahs} and CMS \cite{CMS:2023snh} collaborations 
    {are shown for comparison}.
  }
\EF
%%%%%%%%%%%%%%%%%%%%%%%%%%%%%%%%%%%%%%%%%%%%%%%%%%%%%%%%%%%%%%%%%%%%%%%%%%%%%%%%%%%%

In Fig.~\ref{Fig-UPC1psi}, we present a comparison of our model predictions 
for the rapidity distributions $d\sigma/dy$ of coherent $\Jpsi(1S)$ (left panels) 
and $\psip(2S)$ (right panels) photoproduction in UPC with available data 
covering a broad spectrum of c.m. collision energies,
$\sqrt{s_N} = 200\,\GeV$ (top panels),
$\sqrt{s_N} = 2.76\,\TeV$ (middle panels) 
and $\sqrt{s_N} = 5.02\,\TeV$ (bottom panels).
Our results {were} obtained {for} charmonium wave functions 
generated by two distinct $Q$-$\bar Q$ interaction potentials, POW 
(dotted and dot-dashed lines) and BT (solid and dashed lines).
For the dipole cross sections $\sqq$ we used GBW (solid and dot-dashed lines) 
and KST (dashed and dotted lines) parametrizations.

Fig.~\ref{Fig-UPC1psi} shows a rather good agreement of our predictions based on
the Green function formalism with data at all collision energies.
The best description of data is achieved with the charmonium wave function 
generated by the POW $c$-$\bar c$ potential combined with the GBW model for $\sqq$ 
and/or by the BT $c$-$\bar c$ potential combined with the KST model for $\sqq$. 
The values of $d\sigma/dy$ strongly depend on the shape of the 1S-charmonium wave function. 
Such an observation is consistent with our previous studies 
%--------------------
\cite{Cepila:2019skb}
%--------------------
of quarkonium electroproduction off protons. 
On the other hand, taking into account the latest GBW parametrization of 
$\sqq(r,x)$ from 
%----------------------------
\cite{Golec-Biernat:2017lfv},
%----------------------------
the corresponding calculations lead to magnitudes of $d\sigma/dy$ exhibiting
quite large deviation from values based on the KST model for $\sqq(r,x)$.

Investigation of radially excited charmonia can provide us with 
an additional information about the photoproduction mechanisms in UPC 
due to the nodal structure of their radial wave functions.
Such wave functions are more sensitive to the shape of the $c$-$\bar c$ 
interaction potential compared to 1S-charmonium state
(see Fig.~3 in Ref.~\cite{Cepila:2019skb}).
However, {the} model predictions for {$d\sigma/dy$ and} $\psip(2S)$ production, 
shown in the right panels of Fig.~\ref{Fig-UPC1psi}, exhibit
much weaker sensitivity to the shape of the potential. 
{In fact, there are large differences in $d\sigma/dy$ related to the model for $\sqq$.}
Thus, the measurements of exclusive electroproduction of radially excited charmonia in
nuclear UPC may provide additional constraints on the models for 
{dipole cross section}.

%%%%%%%%%%%%%%%%%%%%%%%%%%%%%%%%%%%%%%%%%%%%%%%%%%%%%%%%%%%%%%%%%%%%%%%%%%%%%%%%%%%%
       %%%%%%%%%%%%%%%%%%%%%%%%%%%%% FIG. 3 %%%%%%%%%%%%%%%%%%%%%%%%%%%%%%%%
%%%%%%%%%%%%%%%%%%%%%%%%%%%%%%%%%%%%%%%%%%%%%%%%%%%%%%%%%%%%%%%%%%%%%%%%%%%%%%%%%%%%
\BF
%%%%%%%%%%%%%%%%
\PSfig{0.36}{dsdyGF_psi1S_5500Pb_BT_KSTr_eikonal_GF.eps}
\hspace*{1.90cm}
\PSfig{0.36}{dsdyGF_psi1S_5500Pb_Pow_GBWnew_eikonal_GF.eps}
\\
\vspace*{0.3cm}
\hspace*{-0.3cm}
\PSfig{0.36}{dsdyGF_psi2S_5500Pb_BT_KSTr_Re+ImV_newGS.eps}
\hspace*{1.7cm}
\PSfig{0.36}{dsdyGF_psi2S_5500Pb_Pow_GBWnew_Re+ImV_newGS.eps}
\\
\vspace*{-0.0cm}
\Caption{
%------------------
\label{Fig-UPC3psi}
%------------------
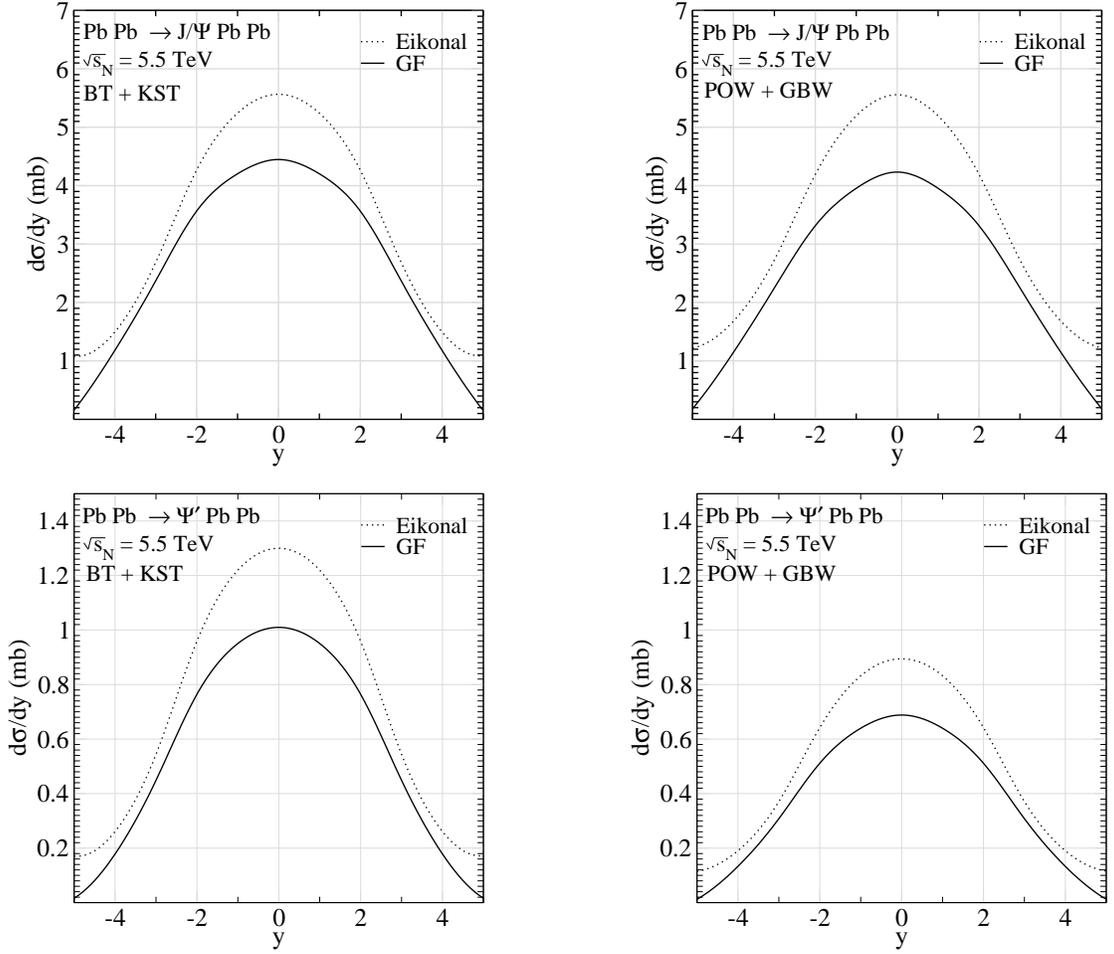
    Manifestations of particular nuclear effects in coherent photoproduction of 
    charmonia in UPC at the LHC collision energy $\sqrt{s_N}= 5.5\,\TeV$. 
    Top and bottom panels correspond to production of 1S and 2S charmonium states, respectively.
    The nuclear cross sections are calculated with charmonium wave functions generated 
    by the BT potential adopting the KST model for the dipole cross section (left panels) 
    and by the POW potential combined with {the} GBW dipole model (right panels).
    The dotted lines represent predictions in the high energy eikonal limit, Eq.~(\ref{coh-lcl}).
    The solid lines are the results of full calculations within {the} Green function 
    formalism including also the gluon shadowing effects.
  }
\EF
%%%%%%%%%%%%%%%%%%%%%%%%%%%%%%%%%%%%%%%%%%%%%%%%%%%%%%%%%%%%%%%%%%%%%%%%%%%%%%%%%%%%

Figure~\ref{Fig-UPC3psi} demonstrates how particular nuclear effects 
modify the rapidity distribution $d\sigma/dy$ at the collision energy 
$\sqrt{s_N}= 5.5\,\TeV$, which is planned to be measured at the LHC.
Here the left and right panels {present} model calculations
including BT and POW $c$-$\bar c$ interaction potential in combination with the
KST and GBW parametrization for $\sqq$, respectively.
Top and bottom panels correspond to model predictions for 1S 
and 2S charmonium states, respectively.
Whereas dashed lines are related to calculations 
{in} the high-energy eikonal approximation, $l_c\gg R_A$ 
(see Eq.~(\ref{coh-lcl})) without corrections to finite coherence length {and GS}, 
the solid lines {were} obtained within {the} Green function formalism
(see Eqs.~(\ref{gf-coh})-(\ref{f1})), accounting also for the leading twist GS effects. 
The differences between the solid and dotted lines are caused by the leading twist GS 
at mid rapidities and by the reduced effects of higher twist quark shadowing 
at forward/backward rapidities.

%%%%%%%%%%%%%%%%%%%%%%%%%%%%%%%%%%%%%%%%%%%%%%%%%%%%%%%%%%%%%%%
%\textbf{
Following discussion in Sec. V.H. of Ref.~\cite{Cepila:2019skb} 
we avoided the skewness
effect in all calculations
due to an absence of
the exact analytical expression for 
the correction factor $R_g(\Lambda)$ in the literature.
The approximate relation for $R_g(\Lambda)$ from Ref.~\cite{Shuvaev:1999ce}
is based on several assumptions,  {such as} the strong inequalities
$x^{\,\prime}\ll x\ll 1$\footnote{Variables $x$ and $x^{\,\prime}$ are the LF
fractions of the proton momentum carried by
the gluons attached to the $Q\bar Q$ fluctuation of the photon.}
and the specific power-law form of the diagonal gluon density of the
target.
This does not allow its application for
 {all} kinematic regions covered by recent UPC experiments.
Nevertheless, the formal implementation of 
the skewness correction from Ref.~\cite{Shuvaev:1999ce}
increases the magnitude of $d\sigma/dy$ by a factor of $\sim 1.4\div 1.5$ for $\Jpsi(1S)$
and by $\sim 1.7\div 2.0$ for $\psip(2S)$ production.
However, it does not lead to any sizeable effect in calculations
of nuclear modification factors $R_A^{coh}$ given by Eq.~(\ref{rcoh}).
%}
%%%%%%%%%%%%%%%%%%%%%%%%%%%%%%%%%%%%%%%%%%%%%%%%%%%%%%%%%%%%%%%%

%%%%%%%%%%%%%%%%%%%%%%%%%%%%%%%%%%%%%%%%%%%%%%%%%%%%%%%%%%%%%%%%%%%%%%%%%%%%%%%%%%%%
       %%%%%%%%%%%%%%%%%%%%%%%%%%%%% FIG. 6 %%%%%%%%%%%%%%%%%%%%%%%%%%%%%%%%
%%%%%%%%%%%%%%%%%%%%%%%%%%%%%%%%%%%%%%%%%%%%%%%%%%%%%%%%%%%%%%%%%%%%%%%%%%%%%%%%%%%%
\BF
%%%%%%%%%%%%%%%%
\hspace*{-0.3cm}
\PSfig{0.36}{dsdyGF_psi1S_200Au_BT_KSTr_Re_ImV_newGS_compMichal.eps}
\hspace*{1.50cm}
\PSfig{0.36}{dsdyGF_psi2S_200Au_BT_KSTr_Re_ImV_newGS_compMichal.eps}
\\
\vspace*{-0.0cm}
\Caption{
%------------------
\label{Fig-UPC6psi}
%------------------
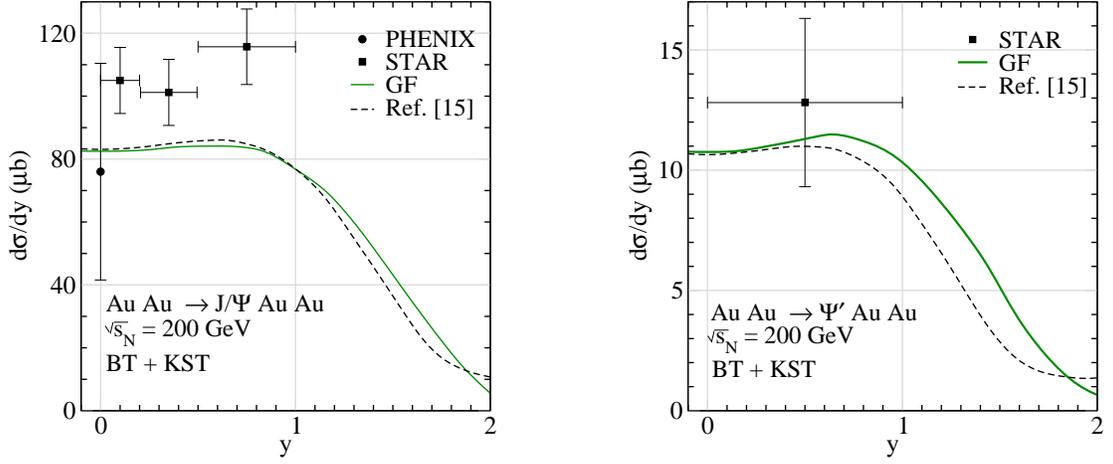
(Color online)
     Rapidity distributions $d\sigma/dy$ of coherent $\Jpsi(1S)$ (left panel) 
     and $\psip(2S)$ (right panel) production in nuclear UPCs at RHIC collision energy
     $\sqrt{s_N}= 200\,\GeV$.
     Present calculations based on path integral technique (solid lines) are compared
     with our previous results from Ref.~\cite{Kopeliovich:2020has} (dashed lines).
     The values of $d\sigma^{coh}/dy$ have been obtained using KST model for $\sqq$
     in combination with BT model for $c$-$\bar c$ interaction potential.
     The data for $d\sigma^{coh}/dy$ from the PHENIX \cite{Afanasiev:2009hy} and 
     STAR \cite{STAR:2023nos,STAR:2023gpk} collaborations {are shown for comparison}.
  }
\EF
%%%%%%%%%%%%%%%%%%%%%%%%%%%%%%%%%%%%%%%%%%%%%%%%%%%%%%%%%%%%%%%%%%%%%%%%%%%%%%%%%%%%

The sophisticated path integral method may significantly improve model
predictions in the kinematic region where $l_c\lsim R_A$.
In Fig.~\ref{Fig-UPC6psi}, we present a comparison of 
{our previous results from Ref.~\cite{Kopeliovich:2020has} with calculations
based on the Green function formalism} at the RHIC collision energy.
For 1S charmonium production they differ by about $10\div 20\%$ at forward/backward 
rapidities.
However, for the coherent production of $\psip(2S)$, such a comparison exhibits 
larger difference, {by $30\div 40\%$,} due to a nodal structure of the corresponding
charmonium wave function and larger averaged size, $\la r^2_{\psip}\ra \textgreater \la r^2_{\Jpsi}\ra$
%---------------------
\cite{Hufner:2000jb}.
%---------------------
%
We predict similar difference between the present and previous results
from Ref.~\cite{Kopeliovich:2020has} also in the LHC kinematic region 
related to values of $y$ {($|y| \sim 4 \div 5$)} that are 
effective for manifestation of reduced effects of quantum coherence.

%%%%%%%%%%%%%%%%%%%%%%%%%%%%%%%%%%%%%%%%%%%%%%%%%%%%%%%%%%%%%%%%%%%%%%%%%%%%%%%%%%%%
       %%%%%%%%%%%%%%%%%%%%%%%%%%%%% FIG. 4 %%%%%%%%%%%%%%%%%%%%%%%%%%%%%%%%
%%%%%%%%%%%%%%%%%%%%%%%%%%%%%%%%%%%%%%%%%%%%%%%%%%%%%%%%%%%%%%%%%%%%%%%%%%%%%%%%%%%%
\BF
%%%%%%%%%%%%%%%%
\hspace*{-0.3cm}
\PSfig{0.36}{RcohGF_psi1S_Au_Pow_GBWnew_1.eps}
\hspace*{1.50cm}
\PSfig{0.36}{RcohGF_psi2S_Au_Pow_GBWnew.eps}
\\
\vspace*{-0.0cm}
\Caption{
%------------------
\label{Fig-UPC4psi}
%------------------
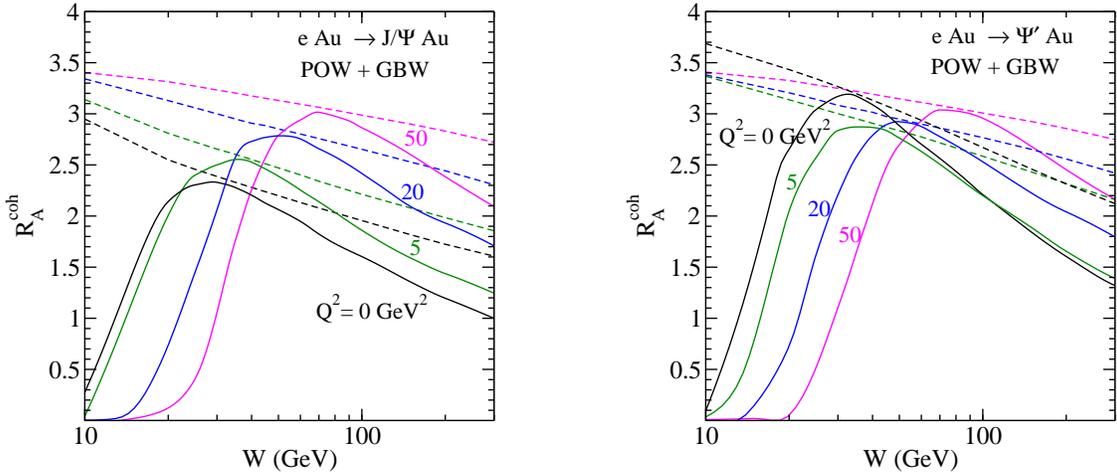
(Color online)
    Ratios $R_A^{coh}$ for the $\Jpsi(1S)$ (left panel) and $\psip(2S)$ (right panel)
    coherent production on the gold target as function of c.m. energy $W$ at 
    several fixed values of the photon virtuality $Q^2 = 0$, 5, 20 and 50$\,\GeV^2$.
    Model predictions are obtained with the GBW parameterization of $\sqq$ and
    POW model for the $c$-$\bar c$ interaction potential. 
    The dashed lines are related to calculations in the high energy eikonal limit, 
    the solid curves represent the exact calculations within a rigorous Green function
    formalism including gluon shadowing corrections. 
  }
\EF
%%%%%%%%%%%%%%%%%%%%%%%%%%%%%%%%%%%%%%%%%%%%%%%%%%%%%%%%%%%%%%%%%%%%%%%%%%%%%%%%%%%%

%%%%%%%%%%%%%%%%%%%%%%%%%%%%%%%%%%%%%%%%%%%%%%%%%%%%%%%%%%%%%%%%%%%%%%%%%%%%%%%%%%%%
       %%%%%%%%%%%%%%%%%%%%%%%%%%%%% FIG. 5 %%%%%%%%%%%%%%%%%%%%%%%%%%%%%%%%
%%%%%%%%%%%%%%%%%%%%%%%%%%%%%%%%%%%%%%%%%%%%%%%%%%%%%%%%%%%%%%%%%%%%%%%%%%%%%%%%%%%%
\BF
%%%%%%%%%%%%%%%%
\hspace*{-0.3cm}
\PSfig{0.36}{RcohGF_Q2+MV2_psi1S_Au_Pow_GBWnew_new.eps}
\hspace*{1.50cm}
\PSfig{0.36}{RcohGF_Q2+MV2_psi2S_Au_Pow_GBWnew_new.eps}
\\
\vspace*{-0.0cm}
\Caption{
%------------------
\label{Fig-UPC5psi}
%------------------
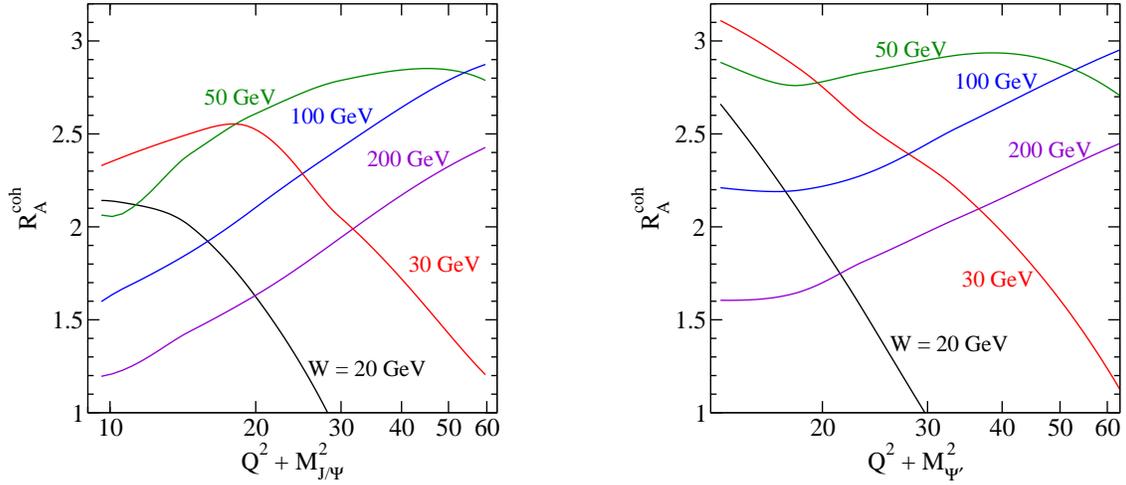
(Color online)
    The same as Fig.~\ref{Fig-UPC4psi} but ratios $R_A^{coh}$ are depicted as function of 
    the variable $Q^2+M_{\Jpsi}^2$ (left panel) and $Q^2+M_{\psip}^2$ (right panel) at 
    several fixed values of the c.m. photon energy $W = 20$, 30, 50, 100 and 200$\,\GeV$.
    The lines represent the exact calculations within a rigorous Green function
    formalism including gluon shadowing corrections.
  }
\EF
%%%%%%%%%%%%%%%%%%%%%%%%%%%%%%%%%%%%%%%%%%%%%%%%%%%%%%%%%%%%%%%%%%%%%%%%%%%%%%%%%%%%

The precise calculations of coherence effects are indispensable for a proper 
description and interpretation of future data on quarkonium electroproduction 
from measurements at planned EIC, {mainly} at large photon virtualities, $Q^2\gg~m_Q^2$,
{where} $l_c\lsim R_A$.
{Therefore,} in Figs.~\ref{Fig-UPC4psi} and \ref{Fig-UPC5psi}, {we demonstrate}
an unambiguous benefit of a rigorous Green function formalism, 
{which correctly incorporates} the higher and leading twist shadowing corrections.
Here we provide predictions for the nucleus-to-nucleon ratios ({\sl{nuclear transparency}}), 
%
%----------------------------------------------------------------------------
\BA
\label{rcoh}
R_A^{coh}(\Jpsi)  = 
\frac{
\sigma_{\gamma^*A\to\Jpsi A}}{A ~\sigma_{\gamma^*N\to\Jpsi N}}
\qquad
\text{and}
\qquad
R_A^{coh}(\psip) = 
\frac{
\sigma_{\gamma^*A\to\psip A}}{A~ \sigma_{\gamma^*N\to\psip N}} \,,
\EA
%---------------------------------------------------
as a function of the c.m. energy $W$ and $Q^2+M_{V}^2$ (V = $\Jpsi(1S)$ and $\psip(2S)$).
The main advantage of investigation of the elastic charmonium electroproduction 
is that CL effects cannot mimic CT. 
{This is because the} contraction of $l_c$ with $Q^2$ (see Eq.~(\ref{lc-hq})) 
{results in an increase 
%\sout{in}} 
{of} suppression of nuclear transparency, 
{which is} an effect opposite to CT.
For this reason, the coherent process, $\gamma^*+A\to\Jpsi (\psip)+A$, 
is convenient for study of particular nuclear effects by future measurements at EIC.
Since manifestations of reduced CL effects are more visible
for heavy nuclei with large radius, we restrict ourselves to predictions 
on the gold target.
{The values of $R_A^{coh}$} were obtained using the charmonium LF wave function 
generated by the POW $c$-$\bar c$ interaction potential and the GBW model for $\sqq$.  
%
%%%%%%%%%%%%%%%%%%%%%%%%%%%%%%%%%%%%%%%%%%%%%%%%%%%%%%%%%%%%%%%%%%%%%%%5
%\textbf
%{
Here, 
due to complexity of calculations within the Green function formalism,
instead of the Dokshitzer-Gribov-Lipatov-Altareli-Parisi (DGLAP) improved GBW saturation
model from Ref.~\cite{Golec-Biernat:2017lfv}, we used a modification of the 
standard GBW model as presented also in Ref.~\cite{Golec-Biernat:2017lfv}
with updated parameters from the fit of deep-inelastic-scattering HERA data
covering higher values of the photon virtuality $Q^2\lsim 50\,\GeV^2$.
%with 
%$Q^2_{max} = 20\,\GeV^2$ and 
%$Q^2_{max} = 50\,\GeV^2$.
%
We expect that corresponding calculations of 
the nucleus-to-nucleon ratio $R_A^{coh}$
at large $Q^2$ 
%using such a modified GBW model 
%are very close to 
will not differ much from
results
based on the DGLAP improved GBW model.
%
%This gives an acceptable accuracy in calculations of $R_A^{coh}$
%at $Q^2\sim 20\div 30\,\GeV^2$.
%
%}
%
Note that the ratio $R_A^{coh}\to A^{1/3}$ at large $Q^2$. 
For this reason the nuclear modification factor $R_A^{coh}$ exceeds one.
%
%%%%%%%%%%%%%%%%%%%%%%%%%%%%%%%%%%%%%%%%%%%%%%%%%%%%%%%%%%%%%%%%%%%%%%%

We predict sizable leading twist corrections, rising with the photon energy,
that can be seen in Fig.~\ref{Fig-UPC4psi} as differences between solid and dashed 
lines at large $W$.
On the other hand, the differences between solid and dashed lines at small and medium
photon energies up to the position of maximal values of $R_A^{coh}$
are caused by reduced CL effects, when $l_c\lsim R_A$ (see Eq.~(\ref{lc-hq})).
Here solid lines clearly demonstrate a diminishing of $R_A^{coh}$ towards 
small photon energies due to suppression of the nuclear coherent cross section
by the nuclear form factor, which is properly included in the Green function formalism.
At large photon energy, 
%\sout{the CL} 
$l_c\gg R_A$ and $R_A^{coh}(W)$ exhibits approximate saturation  
{and then gradual decrease} with $W$ due to the 
%\sout{rising} 
{increasing value of the} dipole cross section with the photon energy. 
The saturation level is higher at larger $Q^2$ which is a clear manifestation of CT effects. 
Our predictions for the onset of reduced coherence effects and gluon shadowing can be tested 
by future experiments at EIC.

Note that model calculations, frequently presented in the literature, 
are usually performed in the high-energy eikonal limit where the coherence 
length $l_c$ exceeds considerably the nuclear size, $l_c\gg R_A$. 
Besides the higher twist corrections, also the gluon shadowing should be included 
{in the LHC kinematic region} {where} $l_c^G = l_c/f_{G}\gg R_A$ with $f_{G}\approx 10$ found in Ref.~\cite{Kopeliovich:2000ra}.
Although such a shadowing correction becomes an important effect at $x\leq 0.01$ 
%------------------------------------------------------------------------------------------------------------
\cite{Kopeliovich:1999am,Kopeliovich:2001xj,Ivanov:2002kc,Nemchik:2002ug,Kopeliovich:2007wx,Kopeliovich:2022jwe}, 
%------------------------------------------------------------------------------------------------------------
it is still missing in the most of recent studies.

In Fig.~\ref{Fig-UPC5psi} we investigate how the $Q^2+M^2_{V}$-behavior of $R_A^{coh}$
changes at different fixed c.m. energies $W$. 
At large $W = $100 and 200$\,\GeV$, the coherence length is sufficiently long 
to neglect its variation with $Q^2+M_V^2$ and to use the “frozen” approximation. 
Then the rise of $R_A^{coh}$ with $Q^2$ is a net manifestation of CT effects.
However, at smaller {photon energies,} $W\leq50\,\GeV$, 
%\sout{the CL} 
$l_c\sim R_A$ and
variation of $l_c$ with $Q^2$ (see Eq.~(\ref{lc-hq})) 
{causes that} the rise of $R_A^{coh}$ gradually slows down and eventually disappears.
%
%\sout{Then} 
At 
%\sout{still smaller} 
$W\leq 20\div 30\,\GeV$, 
%\sout{the CL} 
$l_c\ll R_A$ and the nuclear transparency
$R_A^{coh}$ {decreases rapidly with $Q^2$} due to reduced CL effects.

The right panel of Fig.~\ref{Fig-UPC5psi} shows a non-monotonic $Q^2$ behaviour of $R_A^{coh}$
at small and medium $Q^2$ at the c.m. energy $W=50\,\GeV$,
where the variation of $l_c$ with $Q^2$ does not play any role.
The node effect causes the rise 
of the overlap of the $c\bar c$ state and the 
$\psip(2S)$ wave function {with $Q^2$} in the numerator and denominator of $R_A^{coh}$.
The nuclear medium filters out large size $c\bar c$ configurations 
with larger absorption cross section.
Consequently, the color filtering effect is stronger at smaller $Q^2$ and thus
the $Q^2$ squeezing of the transverse size of a $c\bar c$ wave packet
is less pronounced in the numerator than in the denominator of the ratio $R_A^{coh}$.
As a result, the ratio $R_A^{coh}(Q^2)$ for the coherent production of $\psip$ can
{decrease} at small $Q^2$. 
This may explain also a non-monotonic $Q^2$ dependence of an approximate saturation levels (maxima)
of $R_A^{coh}(W)$ at small $Q^2$ in coherent production of $\psip(2S)$, 
as is demonstrated in the right panel of Fig.~\ref{Fig-UPC4psi}.
At high $Q^2$, however, the size of
the $c\bar c$ fluctuation is so small that color filtering does not play any role.
Then the nuclear medium becomes more transparent 
for smaller $c\bar c$ sizes and may eventually cause a rise 
{of $R_A^{coh}(Q^2)$ with $Q^2$}.
If the CL still exceeds considerably the nuclear radius, such a rise
of $R_A^{coh}(Q^2)$ is a net manifestation of CT effects
(see a rise of $R_A^{coh}(Q^2)$ at $W=50, 100$ and $200\,\GeV$
at medium and large $Q^2$ in the right panel of Fig.~\ref{Fig-UPC5psi}).

%
%
%
%==================================================
\section{Conclusions}
\label{conclusions}
%==================================================
%
%
%

In this paper we present for the first time the comprehensive and uniform
quantum-mechanical description of heavy quarkonium 
production in heavy-ion ultra-peripheral and electron-ion collisions
based on a rigorous Green function formalism, revising thus our previous results from 
%--------------------------------
Ref.~\cite{Kopeliovich:2020has}.
%--------------------------------
%
Here we summarize the following main observations:

\begin{itemize}

\item 
The coherence length $l_c$ for the lowest $|Q\bar Q\ra$ Fock state of the photon exceeds
considerably the nuclear size in UPC at RHIC and the LHC at mid rapidities. 
This enables us to adopt the high-energy eikonal approximation for nuclear effects 
given by Eq.~(\ref{coh-lcl}).
Such an approximation can be reached naturally also within the Green
function formalism in the limit of large photon energies as was demonstrated
in Sec.\ref{gf}.  
The corresponding shadowing correction represents the higher twist effect,
which is small for heavy quarkonia since {it} vanishes with heavy quark mass as $1/m_Q^2$.

\item 
At forward and/or backward rapidities of UPC kinematics at RHIC and the LHC, 
the coherence length $l_c\lsim R_A$ and 
the eikonalization of $\sqq(r,s)$ cannot be applied anymore for calculations 
of nuclear cross sections.
Here we performed for the first time a rigorous and uniform path-integral-technique description.
It does not require to calculate the reduced coherence effects via effective correction 
factors as was done in 
%--------------------------------
Ref.~\cite{Kopeliovich:2020has}. 
%--------------------------------
%
Moreover, such formalism can be applied without restrictions for the coherence length $l_c$,
for arbitrary realistic $Q$-$\bar Q$ interaction potential, as well as for any phenomenological
model for the dipole cross section and/or impact-parameter-dependent partial dipole amplitude.
This represents a substantial improvement of our work 
%--------------------------
\cite{Kopeliovich:2020has},
%--------------------------
modifying our previous results of rapidity distributions $d\sigma/dy$ at small photon energies.
%(see Figs.~\ref{Fig-UPC3psi} and \ref{Fig-UPC6psi}).
%
For 1S charmonium production they differ by about $10\div 20\%$ from our present results. However, for $\psip(2S)$ production they exhibit larger difference, by $30\div 40\%$, due to the nodal structure of the 2S charmonium wave function.

\item 
We have proposed a procedure how to obtain the real part of potentials 
${\mathcal Re} V_{Q\bar Q}(z,\vec r,\alpha)$ in the LF frame 
from various realistic models describing the $Q$-$\bar Q$ interaction in the rest frame.
Such potentials lead to a correct shape of quarkonium wave functions in the LF frame. This is unavoidable for a proper solution of the Schr\"odinger equation 
(\ref{schroedinger}) for the Green function.

\item
The higher Fock components of the photon containing gluons lead to 
gluon shadowing corrections with corresponding coherence length, 
which is much shorter compared to the lowest $|Q\bar Q\ra$ Fock state.
The $|Q\bar QG\ra$ Fock state of the photon gives
the dominant contribution to nuclear shadowing
and represents the leading-twist effect
due to much larger transverse size of the $Q\bar Q$-$G$ dipole 
compared to the small-sized $Q\bar Q$ fluctuation.
The gluon shadowing corrections are effective at mid rapidities 
in the LHC kinematic region, reducing considerably the nuclear cross sections
$d\sigma/dy$
(see Fig.~\ref{Fig-UPC3psi}).

\item
Our model predictions have been calculated employing the
KST and the latest GBW parametrizations for the dipole 
cross section $\sqq(r,s)$, as well as BT and POW
models for the $Q$-$\bar Q$ interaction potential. 
This provided a possibility to estimate a corresponding 
measure of the theoretical uncertainty in our current analysis.

\item
We found a rather good agreement of our revised predictions 
for $d\sigma/dy$ with available data on coherent 
production of $\Jpsi(1S)$ and $\psip(2S)$ in UPC at the energies 
of RHIC and the LHC (see Fig.~\ref{Fig-UPC1psi}). 

\item
We provided also predictions for the nuclear modification factor
$R_A^{coh}$ for coherent charmonium electroproduction as function of 
the c.m. photon energy $W$ and photon virtuality $Q^2$ in the kinematic regions 
covered by the future electron-ion collider at RHIC
(see Figs.~\ref{Fig-UPC4psi} and \ref{Fig-UPC5psi}).
The leading twist effect causes a rise of the suppression of
$R_A^{coh}$ at large photon energies $W$, whereas a vanishing of $R_A^{coh}$  
towards small $W$ is caused by the reduced coherence effects (see 
differences between solid and dashed lines in Fig.~\ref{Fig-UPC4psi}).

\item
Investigation of $R_A^{coh}(Q^2)$ by experiments at EIC is very effective
for study of particular nuclear phenomena, such as CT and CL effects.
Signal of CT may be visible as a rise of nuclear transparency with $Q^2$
at large $W$, where the variation of $l_c$ with $Q^2$ is not important
(see Fig.~\ref{Fig-UPC5psi}).
However, the reduced CL effect causes a decrease in $R_A^{coh}(Q^2)$
with $Q^2$ and so do not mimic CT. 
Thus the onset of the former effect
may lead to a significant modification of $R_A^{coh}(Q^2)$ and
to a complete elimination of any CT signal at medium
energies. Then a strong reduction of $R_A^{coh}$ with $Q^2$ is 
a clear manifestation of the dominance of reduced CL effect
(see lines at $W=20$ and $30\,\GeV$ in Fig.~\ref{Fig-UPC5psi}).

\item 
The precise calculations of the shadowing and absorption effects
in the nuclear medium within a rigorous Green function formalism may be
indispensable for the future conclusive evidence 
of expected gluon saturation effects at large energies.

\end{itemize}

%%%%%%%%%%%%%%%%%%%%%%%%%%%%%%%
\begin{acknowledgments}

The work of J.N. was partially supported 
%by grants ...,  by  the  project  of  the ... 
%and 
by the Slovak Funding Agency, Grant No. 2/0020/22.
%
%The work of J.O. was supported by the project ...
Computational resources were provided by the e-INFRA CZ project (ID:90254),
supported by the Ministry of Education, Youth and Sports of the Czech Republic.

\end{acknowledgments}
%%%%%%%%%%%%%%%%%%%%%%%%%%%%%%%

% ===========================================================

% -----------------------------------------------------------

\end{document}